\documentclass[conference]{IEEEtran}
\IEEEoverridecommandlockouts
\usepackage{cite}
\usepackage{amsmath,amssymb,amsfonts}
\usepackage{algorithmic}
\usepackage{graphicx}
\usepackage{textcomp}
\usepackage{xcolor}
\def\BibTeX{{\rm B\kern-.05em{\sc i\kern-.025em b}\kern-.08em
    T\kern-.1667em\lower.7ex\hbox{E}\kern-.125emX}}

\usepackage{balance} % For balanced columns on the last page
\usepackage{booktabs} % For formal tables

\newtheorem{theorem}{\bf Theorem}
\newtheorem{definition}{\bf Definition}
\newtheorem{lemma}{\bf Lemma}

\makeatletter
\newsavebox{\@brx}
\newcommand{\llangle}[1][]{\savebox{\@brx}{\(\m@th{#1\langle}\)}%
  \mathopen{\copy\@brx\mkern2mu\kern-0.9\wd\@brx\usebox{\@brx}}}
\newcommand{\rrangle}[1][]{\savebox{\@brx}{\(\m@th{#1\rangle}\)}%
  \mathclose{\copy\@brx\mkern2mu\kern-0.9\wd\@brx\usebox{\@brx}}}
\makeatother

\usepackage{algorithm}
\usepackage{epsfig}
\usepackage{amsmath}
\usepackage{amssymb}

\long\def\hide#1{{}}
\def\Section{Sec.}

\def\Relation{Rel.}
\def\Relations{Relations}
\def\Figure{Fig.}

\def\Definition{Def.}

\def\naive{na\"\i ve}

\newlength \smallfigwidth
\newlength \bigfigwidth
\makeatletter
\if@twocolumn
\else
\fi
\makeatother

\begin{document}

\def\Google{Google}
\def\LinkedIn{LinkedIn}

\author{
\IEEEauthorblockN{Ahmed Metwally}
\IEEEauthorblockA{\textit{The Anti-Abuse Team} \\
\textit{{\LinkedIn} Corp.}\\
Mountain View, CA \\
ametwally@linkedin.com}
\and
\IEEEauthorblockN{Chun-Heng Huang}
\IEEEauthorblockA{\textit{The Ad Traffic Quality Team} \\
\textit{{\Google} Inc.}\\
Mountain View, CA \\
chunheng@google.com}
}

\title{Scalable Similarity Joins of Tokenized Strings
\thanks{Part of this work was done while the first author was with {\Google} Inc.
The authors like to thank Theodore Hwa of LinkedIn for his insightful discussions and revising the theoretical foundations.}
}

\maketitle

\begin{abstract}

This work tackles the problem of fuzzy joining of strings that naturally tokenize into meaningful substrings, e.g., full names. Tokenized-string joins have several established applications in the context of data integration and cleaning. This work is primarily motivated by fraud detection, where attackers slightly modify tokenized strings, e.g., names on accounts, to create numerous identities that she can use to defraud service providers, e.g., Google, and LinkedIn.

To detect such attacks, all the accounts are pair-wise compared, and the resulting similar accounts are considered suspicious and are further investigated. Comparing the tokenized-string features of a large number of accounts requires an intuitive tokenized-string distance that can detect subtle edits introduced by an adversary, and a very scalable algorithm. This is not achievable by existing distance measure that are unintuitive, hard to tune, and whose join algorithms are serial and hence unscalable.

We define a novel intuitive distance measure between tokenized strings, \emph{Normalized Setwise Levenshtein Distance} ($NSLD$). To the best of our knowledge, NSLD is the first metric proposed for comparing tokenized strings. We propose a scalable distributed framework, \emph{Tokenized-String Joiner} (TSJ), that adopts existing scalable string-join algorithms as building blocks to perform $NSLD$-joins. We carefully engineer optimizations and approximations that dramatically improve the efficiency of TSJ. The effectiveness of the TSJ framework is evident from the evaluation conducted on tens of millions of tokenized-string names from {\Google} accounts. The superiority of the tokenized-string-specific TSJ framework over the general-purpose metric-spaces joining algorithms has been established.
\end{abstract}

\section{Introduction}
\label{sec:intro}

Today, a large number of Internet services are provided to the public, including web-search (e.g., Google Search), online maps (e.g., Google Maps), online product reviews (e.g., Amazon Customer Reviews), online social and professional networks (e.g., Facebook and LinkedIn), video streaming (e.g., YouTube), and sharing rides and residences (e.g., Uber and Airbnb). The high cost of designing, deploying, and maintaining these services is covered by commissions from Internet advertising (e.g., Google, YouTube and Facebook), commissions from e-commerce transactions (e.g., Amazon), members' subscriptions (e.g., LinkedIn), or claiming a share in the sharing economy (e.g., Uber and Airbnb).

For these providers to thrive, it is crucial for them to guarantee high standards on the quality of the offered services. Google, YouTube and Facebook need to protect the Return-On-Investment of their advertisers by ensuring the ads are \emph{viewed} and \emph{clicked} by real surfers, which respectively reflect real exposure to the market and genuine interest in the advertised products. To maintain the credibility of its product reviews, Amazon should show only reviews from real buyers on the listed products. LinkedIn should ensure the profiles targeted by the recruiters represent real professionals. Uber and Airbnb should ensure ride and residence sharing\hide{ is done in good faith and} jeopardizes neither the safety of the hosts nor the guests.

\hide{Wherever there is money, fraud and abuse follow. }
For these providers to guarantee high-quality services, it is of utmost importance to foster an environment of trust with their users. This entails identifying ill-intentioned users and fraudsters who disguise as legitimate users. Failure to identify abusive users may hurt the company's image and opportunities \cite{V13,I16}. A significant body of recent research has focused on identifying abusive users. Malicious applications on social networks were studied in \cite{RHMF16}. Classifying social networks' accounts based on the novelty of their content was proposed in \cite{W10}. The analysis of the edges in the social network graphs was used for detecting fake accounts \cite{W10,CSYP12,YWWGZD14}, and for mitigating bought online reviews \cite{KKM13}. Limiting the sign-up of fake accounts was studied in \cite{TLSC11}. Detection of account takeover and cloning that are then used by the attackers in malicious activities was studied in \cite{JTJ11,ESKV13,BBMPAAPS14,FJDBG16}. Other research focused on identifying the activity/traffic generated by abusive users. Filtering abusive ads traffic was studied in \cite{MAE05,MEAE08,MP11,DGZ12,SM12,DGZ13}, and fake likes and promotion on social networks was studied in \cite{DFJKS14,IOFDFJKS17}.

Targeting the aforementioned \emph{silo} attacks and their traffic is a first line defense in the arms race between the service providers and the abusers. However, in the online world, an abuser can have numerous identities\footnote{We will use ``Identity'' and ``account'' synonymously.}. Hence, silo-attack defenses do not safeguard against attacks that are carried out by attackers controlling numerous identities \cite{MAEZ07,BMBR11}. These identities may not necessarily be active at the same time. These \emph{multi-identity} attacks can be classified roughly into \emph{serial} and \emph{parallel} attacks, or combinations of both\footnote{In reputation systems, similar notions to those of serial and parallel attacks are known as \emph{Whitewashing} and \emph{Sybil} attacks, respectively \cite{HZN09,ZJZN12}.}.

A serial attacker is one who abuses the service provider using a new account every time her old account is terminated. These attackers gain experience with every account termination, and work on reverse-engineering the silo-attack defenses.

A parallel attacker, on the other hand, is one who controls numerous accounts with the service providers that are active at the same time. The goal is to launch a sophisticated attack that causes significant damage by driving little abuse from each account, such that each individual account stays under the radar level of the silo-attack defenses. Examples of the malicious or criminal activity conducted using \emph{rings} of fake or hijacked accounts include click fraud \cite{S09,B14}, content scraping \cite{S13,G16}, posting fake reviews \cite{C16,K18}, fake video views \cite{S14}, and even fake academic paper reviews \cite{R15}.

Seminal research on multi-identity attacks was conducted in context of click fraud \cite{MAE07}, and email spam \cite{ZKYKYCG09}. The work in \cite{CYYP14,MF12,SPDHRP13,WKWWZZ13,WMP13} clusters the accounts to detect rings of abusive accounts. The recent work in \cite{MPDF15,XFH15} extended this clustering approach using multiple signals simultaneously. The underground market at which abusers buy fake accounts in the thousands was studied in \cite{TMGKP13}. 

%\vspace{-2pt}

\subsection{The Motivating Application}
\label{sec:motivation}

%\vspace{-2pt}

This work is motivated by detecting advertising fraud rings. The online ad industry generated \$88 billion in 2017 in the U.S. alone \cite{iab}. In online advertising, an online-content publisher registers its web sites with the network operator, e.g., {\Google}, to display ads on her sites. A publisher receives revenue to a payment instrument, e.g., bank account, for actions on the displayed ads, e.g., views or clicks by surfers.

Sophisticated fraudsters typically sign up for numerous accounts to spread their revenue and circumvent the silo-attack defenses, which is against the policy of most reputable networks \cite{Google18a}. This motivates clustering publisher accounts for discovering potential click fraud rings \cite{CYYP14,MAE07,MF12,WMP13}. Each publisher account has meta-data, i.e., account attributes. For each account attribute, all the accounts are compared pair-wise. The pairs of accounts that are highly similar are used to form edges in a similarity graph for that attribute, where the graph nodes represent accounts. The graph is clustered. The detected clusters flag potential rings.\hide{ The confidence in the potential rings is boosted by cross validating the clusters across multiple attributes \cite{MPDF15}.}

Some of these attributes are difficult to fake, and are hence excellent candidates for account clustering. One example is the name on the bank account to which the ad-traffic revenue is deposited. Currently, most banks around the world would, if a drastic mismatch is detected in the beneficiary name, not automatically credit the money to the account until manually checked by a bank officer. Providing a fake name to {\Google} would result in the bank rejecting the transfer of the ad-traffic revenue from {\Google} to the attacker's bank account, which defeats the purpose of the attack. \hide{Another attribute is the address on the account. {\Google} mails a pin to the address provided by the account creator. This pin has to be entered by the creator in the account sign-up form, so that {\Google} ensures the authenticity of the account address. Entering a bogus address will result in the account not being verified using the mailed pin.}

Opening bank accounts incur overhead for fraudsters \cite{MDKVS12}. Hence, they do not control an infinite number of bank account holders, and strive to maximize their utilization of the resources they control. A Fraudster would try to use the same resource multiple times by slightly altering its form. 

For instance, a bank account holder whose name is ``Barak Obama'' can open multiple bank accounts, and multiple accounts with a service provider under the slightly-edited names ``Obamma, Boraak H.'' or ``Burak Ubama''. When receiving funds to the right bank account number but to a slightly-edited name, the bank officers would not be alarmed. However, these minor well-crafted edits would circumvent the multi-identify attack defenses of the service provider. If the attack defenses employ {\naive} tokenized-string comparison techniques, they would not identify these accounts with the service provider as a fraud ring since they have different bank account names. \hide{Similarly, the addresses ``1600 Amphitheatre Parkway'' can also be changed to ``Amphitheater Pkwy, 1600'', and postal services would still be able to deliver the mail to the altered address, while the service providers fraud defenses missing these two addresses as duplicates.}

The bank account holder signal\hide{ and the address signal}, among other string signals, warrant devising a very scalable technique for comparing tokenized strings in a fuzzy way. Other abuse-detection applications that benefit from string-comparison include detecting paid reviews, detecting fake comments and harassing messages on social networks, and detecting fake job descriptions on professional networks. Moreover, several well-established applications of data integration and cleaning can benefit from comparing tokenizable strings. Such applications include record joining and deduplication in data warehouses, and comparison shopping search engines \cite{GS12}.

As reviewed in \Section~\ref{sec:related}, existing tokenized-string comparison algorithms are serial, and hence cannot scale to self-joining tens of millions of records. These algorithms employ distance measures that are sensitive to the order of the tokens (FMS \cite{CGGM03}), asymmetric (FMS and AFMS \cite{CGGM03}), or are hard to tune since they require setting multiple independent thresholds (all the measures surveyed and proposed in \cite{CRF03,WLF14}).

%\vspace{-2pt}

\subsection{Our Contributions}
\label{sec:contribution}

%\vspace{-2pt}

Our main contribution can be summarized as follows.

\begin{enumerate}
\item{
In \Section~\ref{sec:formulation}, we formulate the problems of fuzzy joins of tokenized strings. Specifically, we define transformations that capture modifications to tokenized strings. We leverage these transformations to define a novel distance between tokenized strings, \emph{Normalized Setwise Levenshtein Distance} ($NSLD$). $NSLD$ is an intuitive metric that is general enough to be applicable to abuse-detection applications as well as the well-established applications of data integration and cleaning. We prove $NSLD$ is a metric, and hence can be leveraged in all flavors of K-nearest-neighbor queries on metric spaces.
}
\item{
In \Section~\ref{sec:algorithms}, we devise \emph{Tokenized-String Joiner} (TSJ), a scalable $NSLD$-joining framework that adopts existing string-join algorithms as building blocks. We carefully engineer optimizations and approximations that trade efficiency for recall.
}
\item{
TSJ is generalizable to several parallelizing paradigms, such as MPI and OpenMP. We report the impressive scalability and efficiency of MapReduce-distributed TSJ on tens of millions of names on {\Google} accounts in \Section~\ref{sec:evaluation}. We discuss the optimizations and the tradeoffs of the approximations, and demonstrate TSJ's superiority over state-of-the-art generic distance-metric algorithms.
}
\end{enumerate}

\section{Problem Statement and Formulation}
\label{sec:formulation}

The aforementioned applications revolve around tokenized strings, i.e., strings that naturally tokenize into substrings that are meaningful to humans, where slightly editing and shuffling these tokens do not change the interpretation of the string drastically. While it is challenging to mimic the human ability to compare text, a practicable distance/similarity measure to compare tokenized strings is devised. This measure is a metric, and hence can be leveraged in all flavors of K-nearest-neighbor queries on metric spaces, e.g., ~\cite{CGCZ16,ML14,TVPMG16}. Most importantly, it possess the properties that facilitate devising a scalable tokenized-string-joining algorithm\footnote{When detecting fraud rings, the joined sets are one and the same, resulting in a self-join, i.e.,  a pair-wise comparisons of all the tokenized strings.}.

%\vspace{-2pt}

\subsection{Notation and Definition}
\label{sec:notation_definition}

%\vspace{-2pt}

A distance $D(\cdot, \cdot)$ over a set $A$ is a function that maps a pair of elements in $A$ to a non-negative real number. A distance $D(\cdot, \cdot)$ is a metric if the following holds $\forall \alpha, \beta, \gamma \in A$ .
\begin{enumerate}
\item{
Identity: $D(\alpha, \alpha) = 0$,
}
\item{
Symmetry: $D(\alpha, \beta) = D(\beta, \alpha)$, and
}
\item{
Triangle Inequality: $D(\alpha, \beta) + D(\beta, \gamma) \geq D(\alpha, \gamma)$.
}
\end{enumerate}

Let $\Sigma$ be a finite alphabet and $\Sigma^{*}$ be a set of all finite-length strings over $\Sigma$. Denote a string $x \in \Sigma^{*}$ as $x_{1} x_{2} \dots x_{|x|}$, where $|x|$ is the length of the string $x$. A substring of $x$ is $x_{i} x_{i+1} \dots x_{j}$, where $1 \leq i, j \leq |x|$. If $i > j$, the substring is the empty string $\epsilon$, where $| \epsilon | = 0$. A string distance $d(\cdot, \cdot)$ is a distance over $\Sigma^{*}$. A tokenizer $t(\cdot)$ is a function that maps a string $x$ to a finite multiset of strings $x^{t} = \{x^{t1}, x^{t2}, \dots, x^{tm}\}$. We call $x^{t}$ a tokenized string of $x$, and we call $x^{ti}$ a token in $x^{t}$. Denote the number of tokens in $x^{t}$ as $\mathcal{T}(x^{t}) = m$, and the aggregate length of tokens of $x^{t}$ as $\mathcal{L}(x^{t}) = \Sigma_{i} |x^{ti}|$. A simple and commonly used tokenizer splits a string into tokens, i.e., substrings, by using whitespaces as separators\hide{ and discards the separators}. A tokenized-string distance $d^{t}(\cdot, \cdot)$ is a distance over all tokenized strings.

Note that, given $d^{t}(\cdot, \cdot)$ and $t(\cdot)$, we can form a string distance by letting $d(x, y) = d^{t}(t(x), t(y))$. Conversely, we formulate the tokenized-string distances in terms of string distances. Since the tokens of any tokenized string are also strings, string distances can be used to form a tokenized-string distance. More formally, $d^{t}(t(x), t(y)) = d^{t}(x^{t}, y^{t}) = f(d(x^{ti}, y^{tj}) \forall x^{ti} \in x^{t}, y^{tj} \in y^{t})$, for some function $f$.

%\vspace{-2pt}

\subsection{Problem Statement}
\label{sec:problem_statement}

%\vspace{-2pt}

Given two sets of tokenized strings, $R = \{r_{1}^{t}, r_{2}^{t}, \dots, r_{S}^{t}\}$ and $P = \{p_{1}^{t}, p_{2}^{t}, \dots, p_{Q}^{t}\}$, a tokenized-string distance, $d^{t}(\cdot, \cdot)$, and a threshold, $T$, the goal is to find all tokenized string pairs $\langle r_{s}^{t}, p_{q}^{t} \rangle$, s.t.  $r_{s}^{t} \in R$, $p_{q}^{t} \in P$ and $d^{t}(r_{s}^{t}, p_{q}^{t}) \leq T$.

The problem can also be expressed in terms of similarity. Given a conversion scheme, $\lambda$, from distance to similarity, the goal would be to find all tokenized string pairs whose similarity is at least $\lambda(T)$. Several such schemes are commonly used, e.g., $\lambda(T) = 1 - T$, $\lambda(T) = 1 / (1 + T)$ or $\lambda(T) = e^{- T}$.

%\vspace{-2pt}

\subsection{String Distances}
\label{sec:distances}

%\vspace{-2pt}

\hide{
Tokens, being strings, are compared using string distances.
}
% Add description on several distance candidates.
% Need to mentioned edit distance, Jaro Distance, Jaro-Winkler Distance, ...
% State the reason why we talk about the details of the following distance...

\subsubsection{Levenshtein Distance (LD)}
\label{sec:ld}

We borrow the Levenshtein Distance first introduced in~\cite{L66}.

\begin{definition}
Denote $\langle a \to b\rangle$ as a character-level edit operation transforming a string $a$ to a string $b$, where $|a|, |b|$ are $0$ or $1$. Three character-level edit operations are used.

\begin{enumerate}
\item{
\emph{Insertion} ($|a|=0, |b|=1$)
}
\item{
\emph{Deletion} ($|a|=1, |b|=0$)
}
\item{
\emph{Substitution} ($|a|=1, |b|=1$)
}
\end{enumerate}

Given two strings $x, y \in \Sigma^{*}$, $LD(x, y)$ is the minimum number of character-level edit operations (Insertion, Deletion and Substitution) that transform $x$ to $y$.
\end{definition}
\begin{lemma}
$LD(\cdot, \cdot)$ is a metric.
\end{lemma}

$LD$ does not consider the string length or the number of matched characters, which biases the distance to be smaller when comparing relatively short strings. For example, the distance $LD(\mbox{``Thomson''}$, $\mbox{``Thompson''})$ $= 1$, while $LD(\mbox{``Alex''}, \mbox{``Alexa''}) = 1$. While both pairs have the same $LD$, humans may consider the former pair more similar. This motivates normalizing the Levenshtein Distance.

\subsubsection{Normalized Levenshtein Distance (NLD)}
\label{sec:nld}

% Several normalization methods have been proposed.
We borrow the Normalized Levenshtein Distance proposed in~\cite{LL07}.
\begin{definition}
\label{definition:NLD}
For any pair $x, y \in \Sigma^{*}$, $NLD(x, y) = \frac{2 \times LD(x, y)}{|x| + |y| + LD(x, y)}$.
\end{definition}

For example, $NLD(\mbox{``Thomson''}, \mbox{``Thompson''}) = \frac{2 \times 1}{7 + 8 + 1} = \frac{1}{8}$, while $NLD(\mbox{``Alex''}, \mbox{``Alexa''}) = \frac{2 \times 1}{4 + 5 + 1} = \frac{1}{5}$. This motivates normalization makes $NLD$ more intuitive than $LD$.

\begin{lemma}
\label{lemma:NLD_range}
$NLD(\cdot, \cdot) \in [0, 1]$.
\end{lemma}

\begin{theorem}
\label{theorem:NLD_metric}
$NLD(\cdot, \cdot)$ is a metric.
\end{theorem}

The proof of Lemma~\ref{lemma:NLD_range} is trivial and that of Theorem~\ref{theorem:NLD_metric} is in~\cite{LL07}. We introduce the following lemma.

\begin{lemma}
\label{lemma:NLD_threshold}
Assuming $|y| \geq |x|$, the following relationship holds $1 - \frac{|x|}{|y|} \leq NLD(x, y) \leq \frac{2}{\frac{|x|}{|y|} + 2}$.
\end{lemma}

%\vspace{-2pt}

\subsection{Tokenized-String Distances}
\label{sec:proposed_distances}

%\vspace{-2pt}

The straightforward way to define a distance between two tokenized strings is to use an existing similarity measure, e.g. Ruzicka, cosine, and Dice~\cite{C07}, to measure the similarity between their token multisets. This proposal is too rigid when considering token edits. A token that belongs to two strings will not be counted as common if it is slightly edited, leading to biasing the distance measures. We propose tokenized-string distances that accommodate token edits\footnote{Other non-metric distances~\cite{CRF03,CGGM03,WLF14} are reviewed in \Section~\ref{sec:related}.}.

\subsubsection{Setwise Levenshtein Distance (SLD)}
\label{sec:sld}

We propose two set-level edit operations, and use them, along with character-level edit operations to devise $SLD$.

\begin{definition}
\label{definition:SLD}
Denote $\langle a^{t} \to b^{t}\rangle$ as a set-level edit operation transforming a tokenized string $a^{t}$ to a tokenized string $b^{t}$. Two set-level edit operations are defined.

\begin{enumerate}
\item{
\emph{AddEmptyToken} $\langle a^{t} \to a^{t} \cup \{\epsilon\}\rangle$
}
\item{
\emph{RemoveEmptyToken} $\langle a^{t} \to a^{t} \setminus \{\epsilon\}\rangle$
}
\end{enumerate}

Given two tokenized strings $x^{t}, y^{t}$, $SLD(x^{t}, y^{t})$ is the minimum number of character-level edit operations (Insertion, Deletion and Substitution) performed on tokens with additionally any number of set-level edit operations (AddEmptyToken and RemoveEmptyToken) that transform $x^{t}$ to $y^{t}$.
\end{definition}

For example, if $x^{t} = \{ \mbox{``chan''}$, $\mbox{``kalan''} \}$, $y^{t} = \{ \mbox{``chank''}$, $\mbox{``alan''} \}$, and $z^{t} = \{ \mbox{``alan''} \}$. $SLD(x^{t}, y^{t})$ $= 2$ by editing ``chan'' to ``chank'' and ``kalan'' to ``alan''. $SLD(x^{t}, z^{t})$ $= 5$ by editing ``kalan'' to ``alan'', ``chan'' to $\epsilon$, and removing the $\epsilon$.

\begin{lemma}
\label{lemma:SLD_metric}
$SLD(\cdot, \cdot)$ is a metric.
\end{lemma}

\subsubsection{Normalized Setwise Levenshtein Distance (NSLD)}
\label{sec:nsld}

$NSLD$ can be defined in terms of $SLD$ as follows.

\begin{definition}
\label{definition:NSLD}
Given two tokenized strings $x^{t}, y^{t}$, \\$NSLD(x^{t}, y^{t}) = \frac{2 \times SLD(x^{t}, y^{t})}{\mathcal{L}(x^{t}) + \mathcal{L}(y^{t}) + SLD(x^{t}, y^{t})}$.
\end{definition}

For example, if $x^{t} =$ $\{ \mbox{``chan''}$, $\mbox{``kalan''} \}$ and $y^{t} = \{ \mbox{``chank''}$, $\mbox{``alan''} \}$. $NSLD(x^{t}, y^{t}) = \frac{2 \times 2}{9 + 9 + 2} = 0.2$. This motivates normalization makes $NSLD$ more intuitive than $SLD$.

\begin{lemma}
\label{lemma:NSLD_threshold_1}
$NSLD(\cdot, \cdot) \in [0, 1]$.
\end{lemma}

\begin{lemma}
\label{lemma:NSLD_threshold_2}
Assuming $\mathcal{L}(y^{t}) \geq \mathcal{L}(x^{t})$, the following relationship holds $1 - \frac{\mathcal{L}(x^{t})}{\mathcal{L}(y^{t})} \leq NSLD(x^{t}, y^{t}) \leq \frac{2}{\frac{\mathcal{L}(x^{t})}{\mathcal{L}(y^{t})} + 2}$.
\end{lemma}

\begin{theorem}
\label{theorem:NSLD_metric}
$NSLD(\cdot, \cdot)$ is a metric.
\end{theorem}

By proving $NSLD$ is a metric, it can be leveraged in all flavors of K-nearest-neighbor queries on metric spaces, e.g., \cite{CGCZ16,ML14,TVPMG16}. Moreover, the existing algorithms for distributed joining on general metric spaces, e.g., \cite{SRT12,WMP13,DHC14,LSW16}, can be leveraged for performing $NSLD$-joins. However, these techniques are not optimized for (tokenized) strings. This motivates developing a specialized framework, Tokenized-String Joiner (TSJ). TSJ leverages existing distributed $LD$-joining algorithms as building blocks, as discussed in \Section~\ref{sec:algorithms}.\hide{ Comparing TSJ with general-purpose metric-spaces joining algorithms is among our future work.}

\section{Joining Tokenized Strings}
\label{sec:algorithms}
The Tokenized-String Joiner (TSJ) follows a generate-filter-verify paradigm for $NSLD$-joins of tokenized strings. This section discusses how to develop TSJ using MapReduce. 

%\vspace{-2pt}

\subsection{The MapReduce Framework}
\label{sec:mr_framework}

%\vspace{-2pt}

MapReduce~\cite{DG04} has become the \textit{de facto} framework for scalable data processing in shared-nothing clusters, since it offers high scalability and built-in fault tolerance. The computation is expressed in terms of two functions, \emph{map} and \emph{reduce}.

%\vspace{-10pt}

\begin{flalign*}
%\vspace{-15pt}
& \textbf{map}: \langle key_1, value_1 \rangle \rightarrow [\langle key_2, value_2 \rangle] &\\
& \textbf{reduce} : \langle key_2, [value_2] \rangle \rightarrow  [value_3] &
%\vspace{-15pt}
\end{flalign*}

Each record in the input dataset is represented as a tuple $\langle key_1, value_1 \rangle$. The input dataset is distributed among the \emph{mappers} that execute the map functionality. Each mapper applies the map function on each input record to produce a list on the form $[\langle key_2, value_2 \rangle]$, where $[.]$ represents a list. Then, the \emph{shufflers} group the output of the mappers by the key. Next, each \emph{reducer} is fed a tuple on the form $\langle key_2, [value_2] \rangle$, where $[value_2]$, the reduce\_value\_list, contains all the $value_2$'s that were output by any mapper with the same $key_2$ value. Each reducer applies the reduce function on the $\langle key_2, [value_2] \rangle$ tuple to produce a list, $[value_3]$.

\hide{
The framework is depicted in Figure~\ref{fig:map_reduce}.

\begin{figure}[t]
\centering
\includegraphics[width=\smallfigwidth]{figures/map_reduce.eps}
\caption{The MapReduce framework.}
\label{fig:map_reduce}
\end{figure}
}

%\vspace{-2pt}

\subsection{The Generate-Filter-Verify Paradigm}
\label{sec:filters_and_verification}

%\vspace{-2pt}

TSJ follows a generate-filter-verify paradigm for high efficiency. It first generates candidate pairs of tokenized strings. Every generated pair either shares at least one token, or has at least a pair of tokens that are highly similar.

Then, TSJ applies low-cost filters to exclude candidate pairs without loss of correctness. The filters reduce the large number of tokenized string comparisons, and are applied at both the token-level and at the tokenized-string-level. Finally, TSJ calculates the $NSLD$ of the remaining candidate pairs for verification. All the stages are parallelized using MapReduce.

%\vspace{-2pt}

\subsection{Generating Shared-Token Candidate Pairs}
\label{sec:candidate_generation_sharing_tokens}

%\vspace{-2pt}

To generate all pairs of tokenized strings that share at least one token, each tokenized string in $R$ and $P$ is output with all its tokens. Then, the tokenized strings sharing a token are grouped together for further verification. For efficiency, identifiers of the tokenized strings and the tokens are used.

In MapReduce notation, the same map function processes $R$ and $P$ strings, and is defined as $r_{s}^{t} \rightarrow [\langle  r_{s}^{ti}, r_{s}^{t} \rangle]$ and $p_{q}^{t} \rightarrow [\langle  p_{q}^{tj}, p_{q}^{t} \rangle]$. The shufflers group by the key, resulting in each token, $z$, being associated with all $r_{s}^{t}$ and $p_{q}^{t}$ tokenized strings containing $z$. Each reducer receives all the tokenized strings $r_{s}^{t}$ and $p_{q}^{t}$ containing a shared token, $z$, and generates the corresponding list of candidates. That is, $\forall r_{s}^{t} | z \in r_{s}^{t}, \forall p_{q}^{t} | z \in p_{q}^{t}$, the reduce function is $\langle z, [r_{s}^{t}] + [p_{q}^{t}] \rangle \rightarrow [\langle r_{s}^{t}, p_{q}^{t} \rangle]$.

%\vspace{-2pt}

\subsection{Generating Similar-Token Candidate Pairs}
\label{sec:candidate_generation_similar_tokens}

%\vspace{-2pt}

TSJ also generates the candidate pairs of tokenized strings that have at least one pair of similar tokens.

\begin{theorem}
\label{theorem:NSLD_to_NLD}
Given two tokenized strings $x^{t}, y^{t}$, and a threshold $T$, s.t. $NSLD(x^{t}, y^{t}) \leq T$. There exists a pair of tokens $\langle x^{ti}, y^{tj}\rangle, x^{ti} \in x^{t}, y^{tj} \in y^{t}$ where $NLD(x^{ti}, y^{tj}) \leq T$.
\end{theorem}

Theorem~\ref{theorem:NSLD_to_NLD} captures the main insight behind the scalability of TSJ. A pair of tokenized strings where all the possible pairs of their tokens have $NLD$ exceeding the threshold $T$ cannot be a candidate for verification. To the best of our knowledge, $NSLD$ is the first distance measure that guarantees two tokenized strings whose distance is below a threshold $T$ have two tokens where a function of LD of the tokens is below $T$. This allows for transforming the join from the tokenized-strings domain to the tokens domain that is more manageable. The number of distinct tokens is typically orders of magnitude smaller than that of distinct tokenized strings.

More formally, $NSLD$-joins of tokenized strings can be reduced to the problem of $NLD$-joins of tokens. To formalize this reduction, we define the token space of a set of tokenized strings to be the set of all the tokens of all of the tokenized strings in the set. Given two sets of tokenized strings, $R = \{r_{1}^{t}, r_{2}^{t}, \dots, r_{S}^{t}\}$ and $P = \{p_{1}^{t}, p_{2}^{t}, \dots, p_{Q}^{t}\}$, define their token spaces as $R^{t} = \{r_{s}^{ti} | r_{s}^{t} \in R, r_{s}^{ti} \in r_{s}^{t}, \forall s, i\}$ and $P^{t} = \{p_{q}^{tj} | p_{q}^{t} \in P, p_{q}^{tj} \in p_{q}^{t}, \forall q, j\}$, respectively. The first phase of generating $NSLD$-candidates for $R$ and $P$ is generating $NLD$-candidates for $R^{t}$ and $P^{t}$. For every pair of tokens, $r_{s}^{ti} \in R^{t}, p_{q}^{tj} \in P^{t}$, such that $NLD(r_{s}^{ti}, p_{q}^{tj}) \leq T$, all the tokenized strings in $R$ and $P$ generating $r_{s}^{ti}$ and $p_{q}^{tj}$, respectively, are $NSLD$-join candidates.

To perform the $NLD$-joins, TSJ employs MassJoin~\cite{DLHWF14}, a MapReduce-distributed version of PassJoin~\cite{LDWF11} that was originally proposed for $LD$-joins. PassJoin intuition is captured in lemma~\ref{lemma:partition_lemma} that is borrowed from~\cite{LDWF11} .

\begin{lemma}
\label{lemma:partition_lemma}
Given two strings $x$, and $y$, and a threshold $U$, where $LD(x, y) \leq U$, partitioning $y$ into any $U + 1$ segments results in at least one segment being a substring of $x$.
\end{lemma}

PassJoin partitions every token $r_{s}^{ti} \in R^{t}$ into its segments, and generates the substrings of every token $p_{q}^{tj} \in P^{t}$. For every matching segment and substring, their generating token pair $\langle r_{s}^{ti}, p_{q}^{tj} \rangle$ is a $LD$-join candidate.
\hide{
Notice that generating the segments from $P^{t}$ and the substrings from $R^{t}$ yields a different set of candidate token pairs. A pair of candidate tokens whose $LD \leq T$ must exist in both sets of token-pair candidates.
}

When generating the segments or the substrings of a token, $z$, the MassJoin mapper outputs $z$ keyed by each of its string chunks (segments or substrings). All tokens sharing the same string chunk are grouped together by the shufflers and are processed by the same reducer. Each reducer outputs all possible candidates pairs of tokens, $x$ and $y$, from $R^{t}$ and $P^{t}$, respectively. The candidates are de-duplicaated, and the $LD$-similar pairs of tokens are produced. MassJoin augments the mapper output key by metadata to reduce candidate pairs, and whenever possible, uses unique ids of chunks and tokens. We next explain how to adopt the $LD$-MassJoin for $NLD$-joins.

\begin{lemma}
\label{lemma:NLD_to_LD_threshold}
Given two strings $x$, and $y$, and a threshold $T$, s.t. $NLD(x, y) \leq T$. If $|x| \leq |y|$ then $LD(x, y) \leq \lfloor \frac{2 \times T \times |y|}{2-T} \rfloor$.  If $|x| > |y|$, then $LD(x, y) \leq \lfloor \frac{T \times |y|}{1-T}  \rfloor$.
\end{lemma}

\begin{lemma}
\label{lemma:lowerbound_for_x}
Given two strings $x$, and $y$, and a threshold $T$, s.t. $NLD(x, y) \leq T$. If $|x| \leq |y|$, then $\lceil (1 - T) \times |y| \rceil \leq |x|$.
\end{lemma} 

Lemmas~\ref{lemma:partition_lemma} and~\ref{lemma:NLD_to_LD_threshold} establish a lower bound on the number of segments per token. From lemma~\ref{lemma:partition_lemma} any partition scheme is usable. However, an even-partition scheme, where the difference between the shortest and longest generated segments is at most one, reduces the space of string chunks. 

Lemmas~\ref{lemma:NLD_to_LD_threshold} and~\ref{lemma:lowerbound_for_x} establish a condition on the lengths of two tokens to be compared. The \emph{length-condition} $\lceil (1 - T) \times |y| \rceil \leq |x| \leq |y|$ has to hold for a pair of tokens, $x$ and $y$, such that one of which is in $R^{t}$ and the other is in $P^{t}$, and $NLD(x, y) \leq T$.

\hide{Performing the $NLD$-joins on tokens that satisfy the length-condition can be done by dissecting each of $R^{t}$ and $P^{t}$ into token subspaces, such that each token subspace of $R^{t}$ ($P^{t}$) comprise the tokens that have the same length. The token subspaces from $R^{t}$ and $P^{t}$ that satisfy the length-condition are joined using the LD-based MassJoin. To simplify the implementation, the map function of MassJoin was modified. When producing all the substrings of a token, $x$, the output key of the mapper was augmented with $|y|$. The mappers output $x$ with all possible values of $|y|$ for which the length-condition holds. Meanwhile, for every partitioned token, $y$, the mapper augments the key with $|y|$ for the shufflers to be able to perform the join based on the mapper key.}

%\vspace{-2pt}

\subsection{Filtering Candidate Pairs of Tokenized-String}
\label{sec:candidate_filteriing}

%\vspace{-2pt}

When generating candidate pairs of tokenized strings, relying solely on Theorem~\ref{theorem:NSLD_to_NLD} results in a large proportion of spurious candidates, especially when the tokenized strings have numerous tokens. TSJ applies two low-cost filters to effectively prune the candidates before computing their $NSLD$.

\subsubsection{Pruning based on Length}
\label{sec:length_based_filter}

Based on lemma~\ref{lemma:NSLD_threshold_2}, TSJ discards candidate pair of tokenized strings if their aggregate token length ensures their $NSLD$ distance exceeds the threshold, $T$. \hide{The lengths of tokens and tokenized strings that should be filtered are computed once and cached.} The algorithm represents each tokenized string by a unique identifier for efficiency. This identifier is augmented with the length of the tokenized string to prune candidate pairs of identifiers based on the lengths of the tokenized strings.

\subsubsection{Pruning based on Distance Lower Bound}
\label{sec:distance_lower_bound_based_filter}

TSJ can prune candidate pairs by computing a lower bound on the $NSLD$ of any candidate pair by establishing a lower bound on the character-level edit operations for each token of the two tokenized strings. To that end, TSJ augments each tokenized string unique id with a histogram of its token lengths. Given the pair of the token-length histograms of two tokenized strings, and the lengths of the token pairs whose $NLD$ is below $T$, TSJ can compute lower bounds on $SLD$ and $NSLD$ of the pair by computing a lower bound on the character-level edit operations for all the pairs of tokens, whether matched or unmatched. For the matched tokens, the character-level edit operations are already computed during the candidate generation phase. For the unmatched tokens, TSJ computes a lower bound based on the length histograms, and on Lemma~\ref{lemma:histogram_NLD_to_LD_threshold}. 

\begin{lemma}
\label{lemma:histogram_NLD_to_LD_threshold}
Given two strings $x$, and $y$, and a threshold $T$, s.t. $NLD(x, y) > T$. If $|x| \leq |y|$ then $LD(x, y) > \lfloor \frac{T \times |y|}{2 - T} \rfloor$.  If $|x| > |y|$, then $LD(x, y) > \lfloor \frac{2 \times T \times |y|}{2-T}  \rfloor$.
\end{lemma}
 
The pruning algorithm and the proof of its correctness will be discussed in an extended version of the paper.

%\vspace{-2pt}

\subsection{The Final Verification}
\label{sec:verification}

%\vspace{-2pt}

Once the pruned candidates are found, the tokenized-string identifiers are resolved to the tokenized strings, and the final verification is carried by calculating its $SLD$ as below. The $NSLD$ of any pair is trivially computed from its $SLD$.

\subsubsection*{SLD Calculation}
Given two tokenized strings, $x^{t}$, and $y^{t}$, let $x^{t} = \{x^{t1}, x^{t2}, \dots, x^{tm}\}$ and $y^{t} = \{y^{t1}, y^{t2}, \dots, y^{tn}\}$. Calculating $SLD(x^{t}, y^{t})$ entails forming a weighted bigraph $\langle V, E, w\rangle$. Let $k = max(m, n)$. Auxiliary tokenized strings $x^{t \prime} = \{x^{t1}, x^{t2}, \dots, x^{tk}\}$ and $y^{t \prime} = \{y^{t1}, y^{t2}, \dots, y^{tk}\}$ are constructed, where extra tokens are empty strings. Let $X$ be a set of nodes that represents tokens in $x^{t \prime}$ and $Y$ be a set of nodes that represents tokens in $y^{t \prime}$. Define $V = X \cup Y$, $E = \{\langle X_{i}, Y_{j}\rangle | X_{i} \in X, Y_{j} \in Y\}$, and edge weights $w(\langle X_{i}, Y_{j}\rangle) = LD(x^{ti}, y^{tj}), \forall i, j$. The minimum weight perfect matching on this weighted bigraph is computed. This is a manifestation of the assignment problem that can be solved using the Hungarian algorithm. The time complexity for constructing the weighted bigraph is $O(\mathcal{L}(x^{t}) \times \mathcal{L}(y^{t}))$; and the time complexity of the Hungarian algorithm is $O(max(\mathcal{T}(x^{t}), \mathcal{T}(y^{t}))^{3})$. So, the overall time complexity of calculating $SLD(x^{t}, y^{t})$ is $O(\mathcal{L}(x^{t}) \times \mathcal{L}(y^{t}) + max(\mathcal{T}(x^{t}), \mathcal{T}(y^{t}))^3)$.

%\vspace{-2pt}

\subsection{Optimizations and Approximations}
\label{sec:variation}

%\vspace{-2pt}

The section describes optimizations and approximations.

\subsubsection{Self-Joins}
The motivating application focuses on self-join, i.e., $R = P$. Performing self-join allows for a major optimization, skipping symmetric steps in the candidate generation and the pruning steps. When generating  segments, from lemma~\ref{lemma:NLD_to_LD_threshold}, the case where $|x| \leq |y|$ only needs to be considered, yielding fewer segments. When generating Similar-Token candidate pairs, applying the length-condition does not need to consider the cases where $x \in R^{t}$, and $y \in P^{t}$ as well as the cases where $x \in P^{t}$, and $y \in R^{t}$, since $P = R$.

\hide{
Multi-way joins is a generalization where multiple sets of tokenized strings are to be joined. MapReduce-based multi-way joins have been studied \cite{ZCW12} in the context of theta joins. Among our future work is to borrow insights from \cite{ZCW12} into the context of joining tokenized strings, especially when estimating the cost of joining two/multiple sets of tokenized strings (the weights on the edges of the join graph).
}

\subsubsection{High-Frequency Tokens}
Avoiding high-frequency tokens can enhance the quality of comparison results. In the application of comparing full names described in \Section~\ref{sec:motivation}, ``John'' and ``Mary'' are very common tokens, and may result in uninteresting results. Both the Shared-Token and Similar-Token candidate pair generation processes discard tokens that are shared by more than a given maximum number of tokenized strings, $M$. \hide{However, when verifying the unpruned generated candidates, all the original tokens in the tokenized strings are used to compute the exact $NSLD$. }Dropping high-frequency tokens achieves better load balancing between the MapReduce workers.

Dropping high-frequency tokens in a scalable way will be discussed in an extended version of the paper.

\subsubsection{De-Duplicating by Grouping on One String}
During the evaluation of the $NLD$ of tokens or the $NSLD$ of the tokenized strings, duplicate pairs arise. To avoid redundant computation, duplicate candidate pairs are discarded using a MapReduce job. There are two possible strategies. The first one, \emph{grouping-on-both-strings} is to use the MapReduce shuffler to group together instances of the same pair, and make the reducer output this pair exactly once. The second strategy, \emph{grouping-on-one-string}, forms a key-value tuple from each pair, where the key is one string, and the value is the other string. The reducer then de-duplicates the reduce\_value\_list using a hash set. To balance the load among the reducers, for every pair of strings, $\tau$ and $\upsilon$, $\tau$ is used as the key and $\upsilon$ is used as the value if and only if $int(HASH(\tau) < HASH(\upsilon)) = (HASH(\tau) + HASH(\upsilon)) \% 2$, where $HASH$ is a fingerprint function that hashes $\tau$ and $\upsilon$ to an integer. Otherwise, $\upsilon$ becomes the key and $\tau$ becomes the value.

\subsubsection{Exact-Token-Matching Approximation}
When the threshold, $T$, is small, and the tokens are relatively short, a pair of similar tokenized strings tend to share a common token. Then, the candidate-pair generation process can be reduced down to the shared-token strategy. The expensive similar-token strategy can be skipped with minor impact on the recall.

\subsubsection{Greedy-Token-Aligning Approximation}
This approximates the $SLD(x^{t}, y^{t})$ as follows. The edge weights of the token bigraph are computed exactly. However, instead of exactly computing the minimum weight perfect matching using the Hungarian algorithm, a greedy approach is followed. The \emph{greedy-token-aligning} selects the edge with the minimum $LD$ weight, and removes the two corresponding tokens from the bigraph nodes. This is repeated until all the tokens have been matched. This reduces the time complexity to $O(\mathcal{L}(x^{t}) \times \mathcal{L}(y^{t}) + \mathcal{T}(x^{t})\times \mathcal{T}(y^{t})\log{(\mathcal{T}(x^{t})\times \mathcal{T}(y^{t}))})$.

\section{Related Work}
\label{sec:related}

To the best of our knowledge, this is the first work that introduces a distance measure for tokenized strings that is intuitive, metric, and whose join algorithm is scalable. The previous work on tokenized-string similarity, distributed joins of strings, and distributed fuzzy metric joins are reviewed.

The MGJoin algorithm \cite{RLWDCT13} proposed joining tokenized strings by representing each tokenized string as a bag of tokens. This work proposes a serial algorithm, and a MapReduce extension that employ prefix-filtering \cite{GIJKMS01,SK04,CGK06,BMS07,XWLYW11} and inverted indexes. The technique is very similar to \cite{VCL10}, but employs multiple global orders of the tokens. Distributed prefix-filtering-based techniques were shown to be inefficient in \cite{MF12}, as apparent from comparing the speedups against \cite{VCL10} in both \cite{RLWDCT13} and \cite{MF12}. All these \textit{set-based} techniques handle token shuffles, but do not handle token edits.

In the context of answering K-nearest-neighbor queries for data cleaning, Chaudhuri \emph{et al.} introduced Fuzzy Matching Similarity (FMS) \cite{CGGM03}. The user sets penalties for token insertion, deletion, or editing. While being able to handle both token shuffles and edits, FMS had two main drawbacks. First, it is provably not a metric. Second, FMS is sensitive to token order. An approximation, AFMS, was introduced that ignores the token positions. AFMS matches each token in a string to its best matching token in the other string, which may result in multiple tokens from one string matched to the same token in the other string. \hide{AFMS takes the average matching distances across all tokens as the final distance, which has very little similarity, if any, to how humans actually compare text.} Unfortunately, both FMS and AFMS are not symmetric, which poses challenges when using them as tokenized-string similarity measures in other applications. Chaudhuri \emph{et al.} proposed a serial FMS-based query algorithm, FuzzyMatch, to identify the closest K tokenized strings given a query, and devised enhancements for indexing, and caching.

Several tokenized-string comparison approaches are contrasted in \cite{CRF03} that focuses on joining names and records by comparing string. Among the contrasted distance measures are the Jaro-based distances \cite{J95, W99} that emerged from the statistical communities and deals with names as non-tokenized strings. To overcome the non-tokenization, and the set-based distances drawbacks (no shuffles and no edits, respectively), \cite{CRF03}  introduced SoftTfIdf  that computes the Jaccard index of two-tokenized strings while considering the popularity of their tokens. SofTfIdf allows for token edits, by allowing tokens to match with up to some threshold on their Jaro-Winkler (JW) distance. However, SoftTfIdf has multiple drawbacks. The work in \cite{CRF03} does not describe an algorithm to compute the distance. To compare two tokenized strings, two thresholds have to be set by the user, $T1$ on the JW distance between tokens and another threshold, $T2$, on the Jaccard similarity of the tokenized strings. Two tokenized strings are considered $T1$-$T2$-similar if their Jaccard similarity exceed $T2$ given that $T1$ is used as the token matching threshold. Setting two unrelated thresholds impairs the tuning of the join. SoftTfIdf is non-metric, since JW violates the triangle inequality.

The work in \cite{WLF14} is a generalization and an improvement over that in \cite{CRF03}. In addition to Jaccard, the work in \cite{WLF14} adapts more set-based similarity measures \cite{C07}, e.g., Dice, cosine, to tokenized-string joins. Like SoftTfIdf \cite{CRF03}, to compare two tokenized strings, Wang \textit{et al.} require a threshold, $T1$ on token similarity and another threshold, $T2$, on tokenized-string similarity. The algorithm matches two tokens from the two token sets only if their token similarity exceed $T1$. The sum of the similarities of the matched tokens are then used to compute the customized set-based similarity of the two tokenized strings. Like \cite{CGGM03, CRF03}, the similarity measure in \cite{WLF14} depends on two totally unrelated thresholds, which impairs the tuning of the join. Like \cite{CRF03,CGGM03}, the proposed tokenized-string measures are provably non-metric.

Wang \textit{et al.} devised an effective candidate string-pair pruning technique that works for the customized similarity measures in \cite{WLF14}, and is based on prefix-filtering. \hide{In practice, setting $T1 = 0$ allows for the most flexible matching of tokens, but makes the prefix-filtering least effective. }The custom pruning proposed in \cite{WLF14} imposes limitations on adopting new string-joining algorithms. Moreover, the algorithm is serial, i.e., cannot scale out to multiple machines for scalability.

When compared to the existing tokenized-string distance measures, the proposed $NSLD$ is a metric, and naturally uses the tokenized-string threshold, $T$, to prune non-similar tokens (Theorem~\ref{theorem:NSLD_to_NLD}). The TSJ framework uses a state-of-the-art distributed string-joining algorithms to perform the joins, and can hence scale out virtually to any input dataset.

Several string-join algorithms have been proposed, as recently surveyed in \cite{YLDF16}, and their performance was compared in \cite{CRF03,HSM07,CHKSS07}. In our implementation, we used MassJoin \cite{DLHWF14} for finding similar tokens. MassJoin is a distributed version of Pass-Join \cite{LDWF11}, which employs a filter-and-verification framework. In the filter step, Pass-Join generates signatures for strings, such that two similar strings are guaranteed to share at least one signature. This way, Pass-Join prunes pairs of strings that are guaranteed to be dissimilar. The remaining candidate pairs are verified in the verification step. Pass-Join was also generalized such that two similar strings are guaranteed to share $K$ signatures, yielding the PassJoinK algorithm \cite{LYWH14}. The PassJoinK was parallelized, yielding the PassJoinKMR and PassJoinKMRS algorithms \cite{LYWH14}. As noted in \cite{YLDF16}, the MassJoin avoids shortcomings of its competitors by frugally generating candidates, and employing light-weight filters.

Distributed joining in general metric spaces has been the focus of \cite{SRT12,WMP13,DHC14,LSW16}, most of which rediscover or borrow ideas from the work in \cite{JS08}. These efforts (recursively) partition the data records into sub-partitions such that similar records belong to the same partitions, or neighboring partitions in the metric space. These techniques compare only two partitions if their candidate records may yield a joined pair. In theory, these techniques can be employed to compare tokenized strings as well as tokens, since both $NSLD$ and $NLD$ are metrics. However, in preliminary experiments, none of these techniques ran to completion within reasonable time when applied to $NSLD$-joins as detailed in \Section~\ref{sec:evaluation}. \hide{The main reason was each partition was a neighbor to numerous partitions. While it is on our roadmap to evaluate these techniques for $NLD$-joins on tokens within TSJ, we expect MGJoin, MassJoin, PassJoinKMR and PassJoinKMRS that were developed specifically for string comparison to be superior to other general metric comparison algorithms.}

\section{Evaluation}
\label{sec:evaluation}

In this section, we report the results of several experiments that evaluate the scalability and the accuracy of TSJ, and the impact of the proposed optimizations and approximations.

The motivation for this work is detecting rings of accounts using their tokenized strings signals. The experiments were run on the names on {\Google} accounts from a specific region. The number of tokenized strings used is $\mbox{44,382,766}$. \hide{Some accounts have multiple (historical) names. The average number of names per account is $1.1$. }The names were tokenized using whitespaces and punctuation characters. The tokenized strings were self-joined resulting in \mbox{$1.9670 \times 10^{15}$} possible pairs.

The default parameters used for the evaluation runs are as follows. The MapReduce jobs are run on $\mbox{1,000}$ machines ($\mbox{1,000}$ Mappers and $\mbox{1,000}$ Reducers) running Ubuntu Linux. Every machine is allowed only $1$G of memory, $5$G of disk, and $0.5$ cpu. The default thresholds on the pairwise $NSLD$ ($T$), and on popular tokens occurring with multiple tokenized strings ($M$) assume the values $0.1$ and $\mbox{1,000}$, respectively\footnote{These parameters may differ across geo-locations depending on the names popularity. Typically, for each major geo-location, a gradient descent search is performed to set these parameters. At each gradient descent evaluation, a sample of the clusters is evaluated by the operations team of {\Google}, and the rates of true positives and the false positives are computed. The values of $0.1$ and $\mbox{1,000}$ constitute a reasonable starting point for the search. For this dataset, setting $M$ to $\mbox{1,000}$ discarded roughly $1\%$ of the tokens.}.

%\vspace{-2pt}

\subsection{Scalability and Speedup}

%\vspace{-2pt}

\begin{figure}
%\begin{minipage}{\smallfigwidth}
\centering
\epsfig{width = \smallfigwidth, figure = 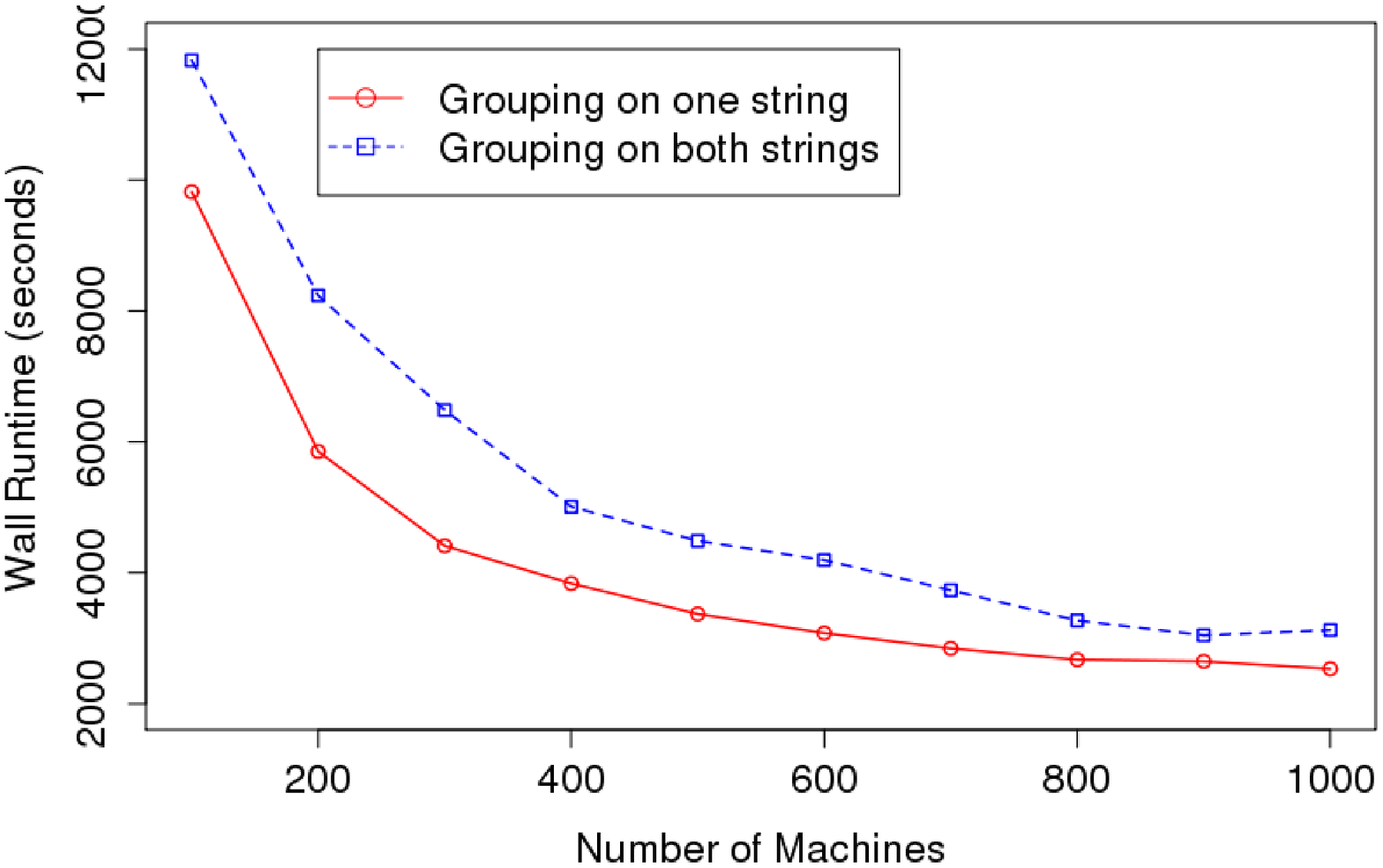}
%\vspace{-10pt}
\caption{Comparing the runtime of Tokenized-String Joiner (TSJ) while varying the MapReduce machines and the Deduping algorithm.}
\label{fig:run_time_vs_num_machines_deduping}
%\vspace{-10pt}
%\end{minipage}
\end{figure}

TSJ was run while varying the number of machines from $100$ to $\mbox{1,000}$\hide{ at a fixed interval of $100$}. Both options for deduping the candidate pairs of strings (grouping-on-one-string and grouping-on-both-strings) were used, and the runtimes are plotted in \Figure~\ref{fig:run_time_vs_num_machines_deduping}. Both deduping options scaled out well. Both achieved a speedup of $3.8$ as the number of machines increased by $10$ folds.

Consistently, grouping-on-one-string was clearly faster than grouping-on-both-strings, achieving a speedup between $13\%$ and $32\%$. This can be attributed to the overhead of instantiating MapReduce workers. The grouping-on-one-string mechanism instantiates a worker for each string, whose work is verifying each of its potentially similar strings. Meanwhile, grouping-on-both-strings instantiates a worker for each candidate pair of strings, whose work is verifying only one pair of strings.

Notice that grouping-on-both-strings achieves better load balancing. In case there exists a small set of strings, each of which is potentially similar to numerous strings, all these candidate pairs would be spread out among multiple workers, which better distributes the load than grouping-on-one-string.

%\vspace{-2pt}

\subsection{Joining by Approximate Matching}

%\vspace{-2pt}

The effect of the approximations in \Section~\ref{sec:variation}, greedy-token-aligning and exact-token-matching, on runtime and accuracy are reported.

\subsubsection{Impact of Approximations on Speedup}

\begin{figure}
%\begin{minipage}{\smallfigwidth}
\centering
\epsfig{width = \smallfigwidth, figure = 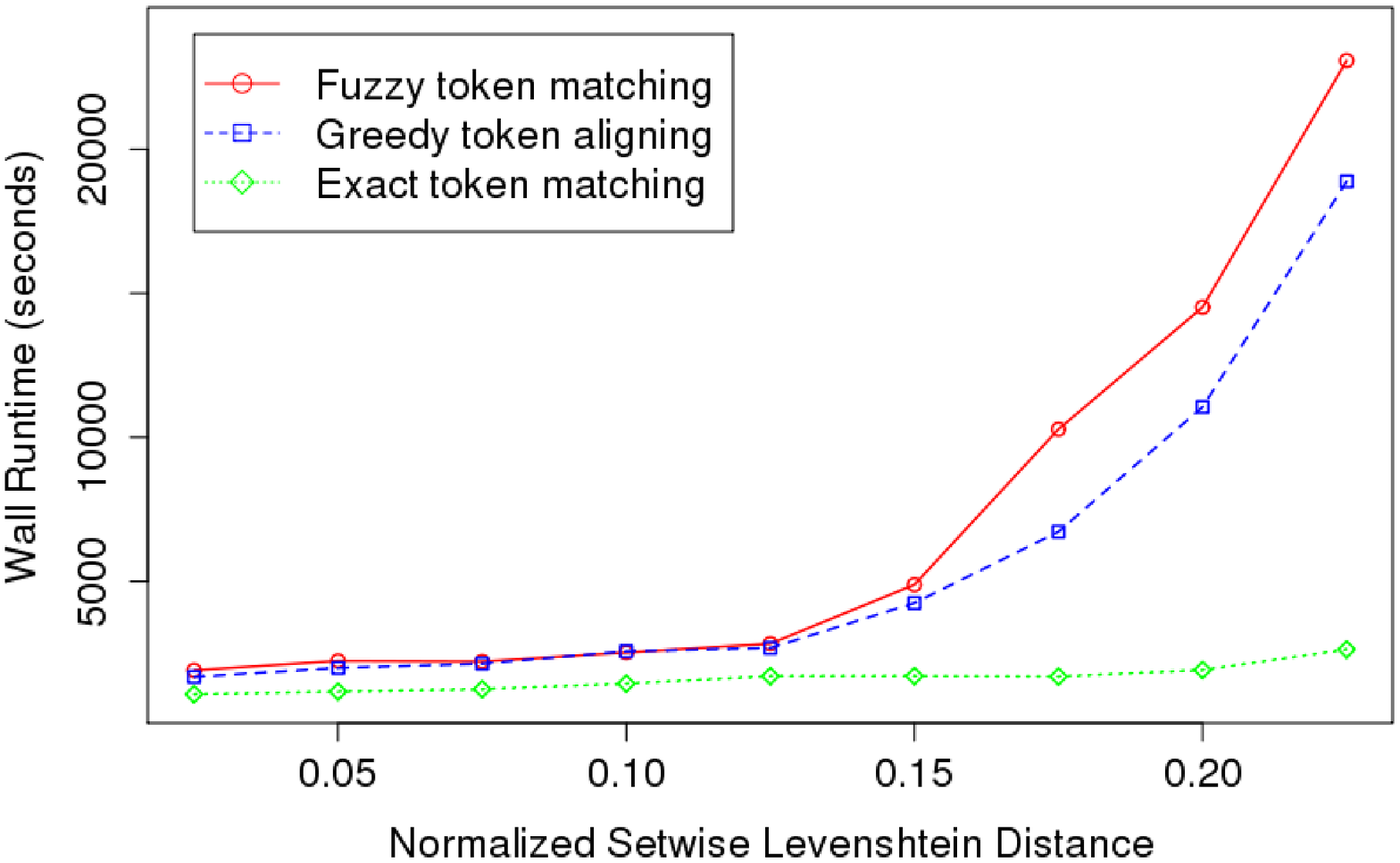}
%\vspace{-10pt}
\caption{Comparing the runtime of Tokenized-String Joiner (TSJ) while varying NSLD and the token matching and aligning algorithms.}
\label{fig:run_time_vs_nsld}
%\end{minipage}
%\vspace{-10pt}
\end{figure}

\begin{figure}
%\begin{minipage}{\smallfigwidth}
\centering
\epsfig{width = \smallfigwidth, figure = 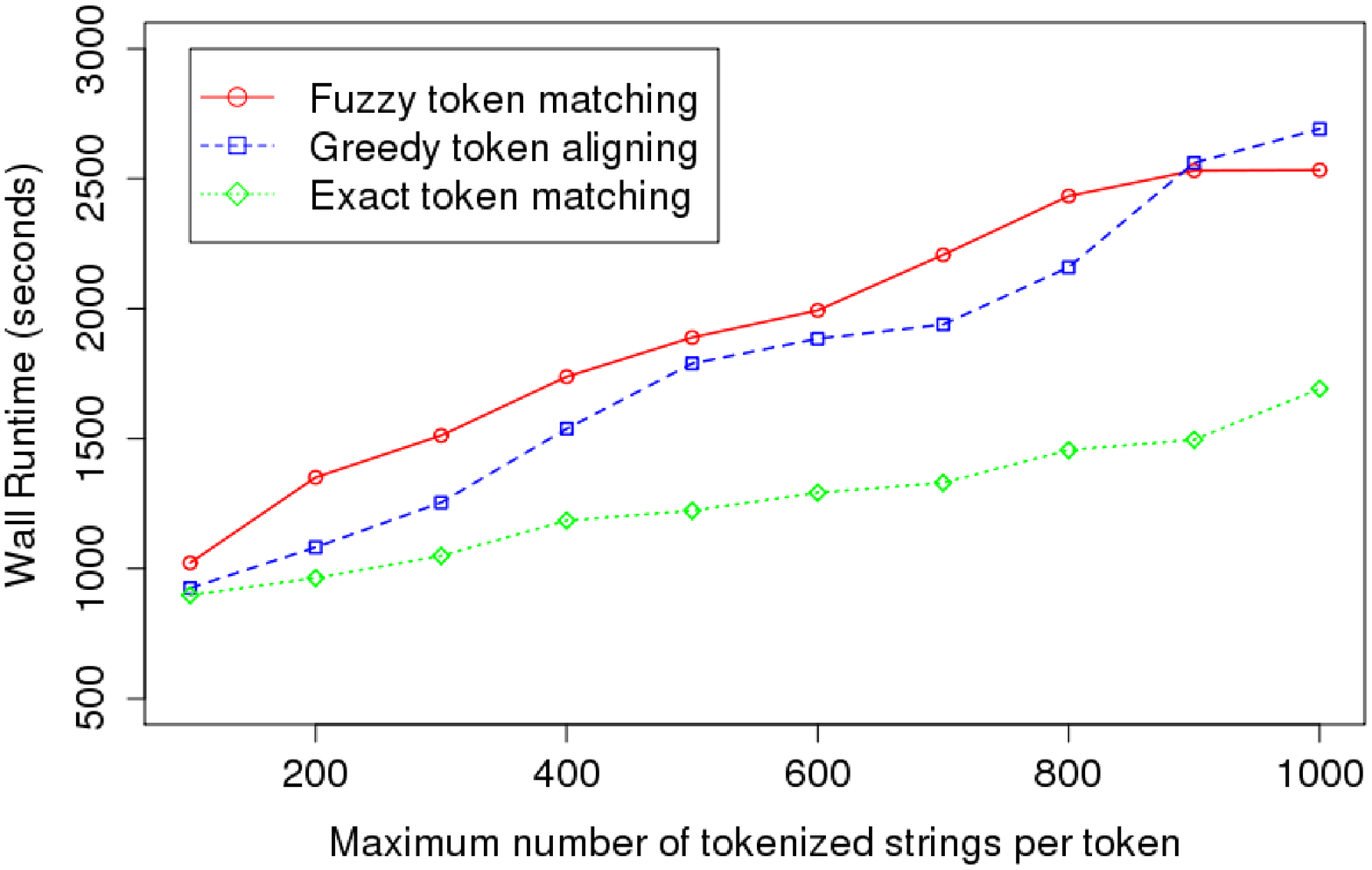}
%\vspace{-10pt}
\caption{Comparing the runtime of Tokenized-String Joiner (TSJ) while varying max-frequency ($M$) and the token matching and aligning algorithms.}
\label{fig:run_time_vs_max_freq}
%\end{minipage}
%\vspace{-5pt}
\end{figure}

All the experiments below are run using grouping-on-one-string. While fuzzy-token-matching produces the correct pairs of similar tokenized strings, it has the longest runtime.

\Figure~\ref{fig:run_time_vs_nsld} shows the runtime while varying $T$ from $0.025$ to $0.225$\hide{ at a fixed interval of $0.025$}. The mean runtime saving of greedy-token-aligning over fuzzy-token-matching is $13\%$, and is more pronounced as $T$ increases. The mean runtime saving of exact-token-matching over fuzzy-token-matching is $60\%$. Moreover, the runtime of exact-token-matching increases only slightly as $T$ increases.

\Figure~\ref{fig:run_time_vs_max_freq} shows the runtime while varying $M$ from $100$ to $\mbox{1,000}$\hide{ at a fixed interval of $100$}. The mean runtime saving of greedy-token-aligning over fuzzy-token-matching is $9\%$, while that of exact-token-matching is $33\%$. The runtime savings of both approximation schemes were fairly stable across the values of $M$.

\subsubsection{Impact of Approximations on Accuracy}

\begin{figure}
%\begin{minipage}{\smallfigwidth}
\centering
\epsfig{width = \smallfigwidth, figure = 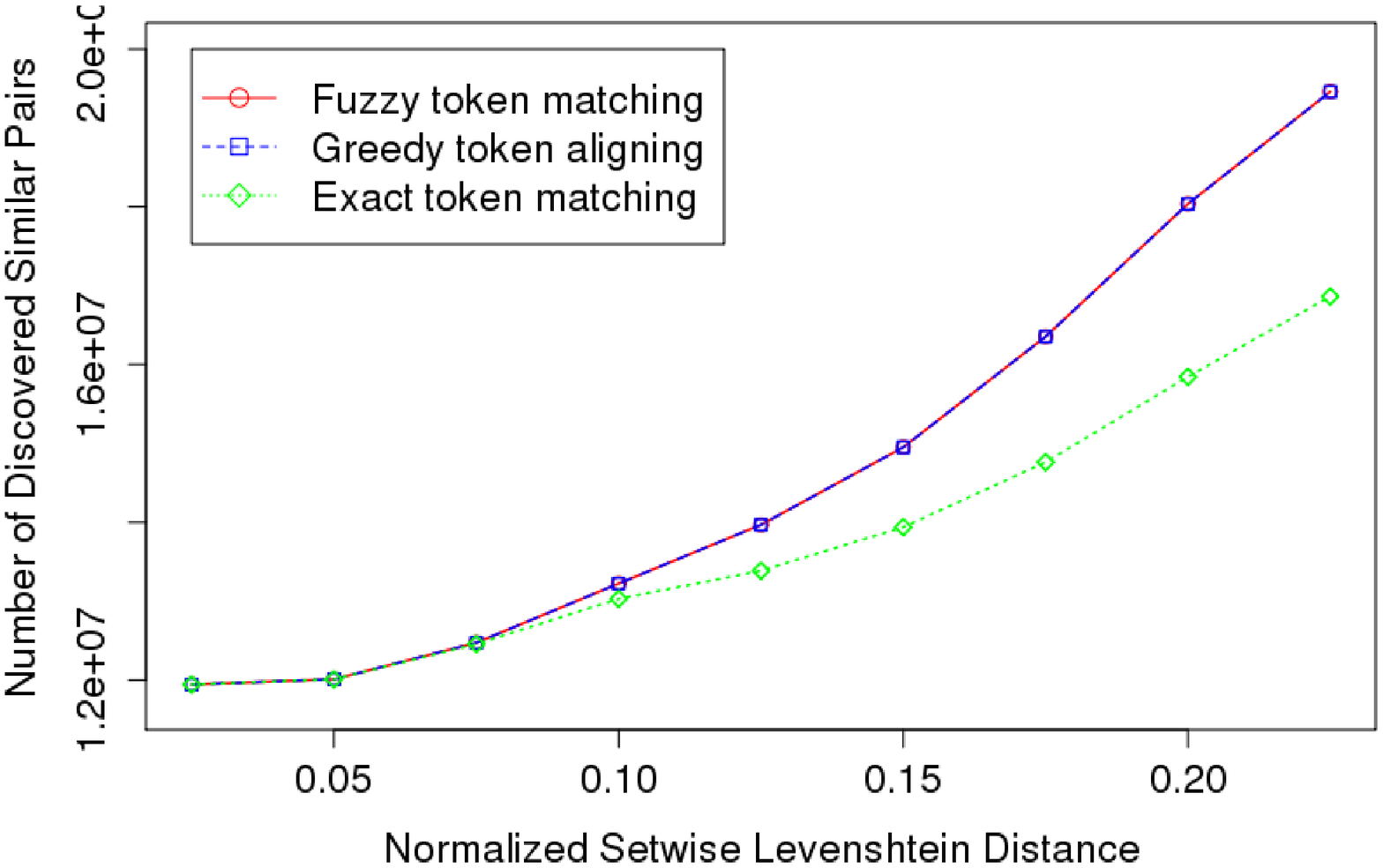}
%\vspace{-10pt}
\caption{Comparing the number of pairs of Tokenized-String Joiner (TSJ) while varying NSLD and the token matching and aligning algorithms.}
\label{fig:num_pairs_vs_nsld}
%\end{minipage}
%\vspace{-10pt}
\end{figure}

\begin{figure}
%\begin{minipage}{\smallfigwidth}
\centering
%\vspace{-15pt}
\epsfig{width = \smallfigwidth, figure = 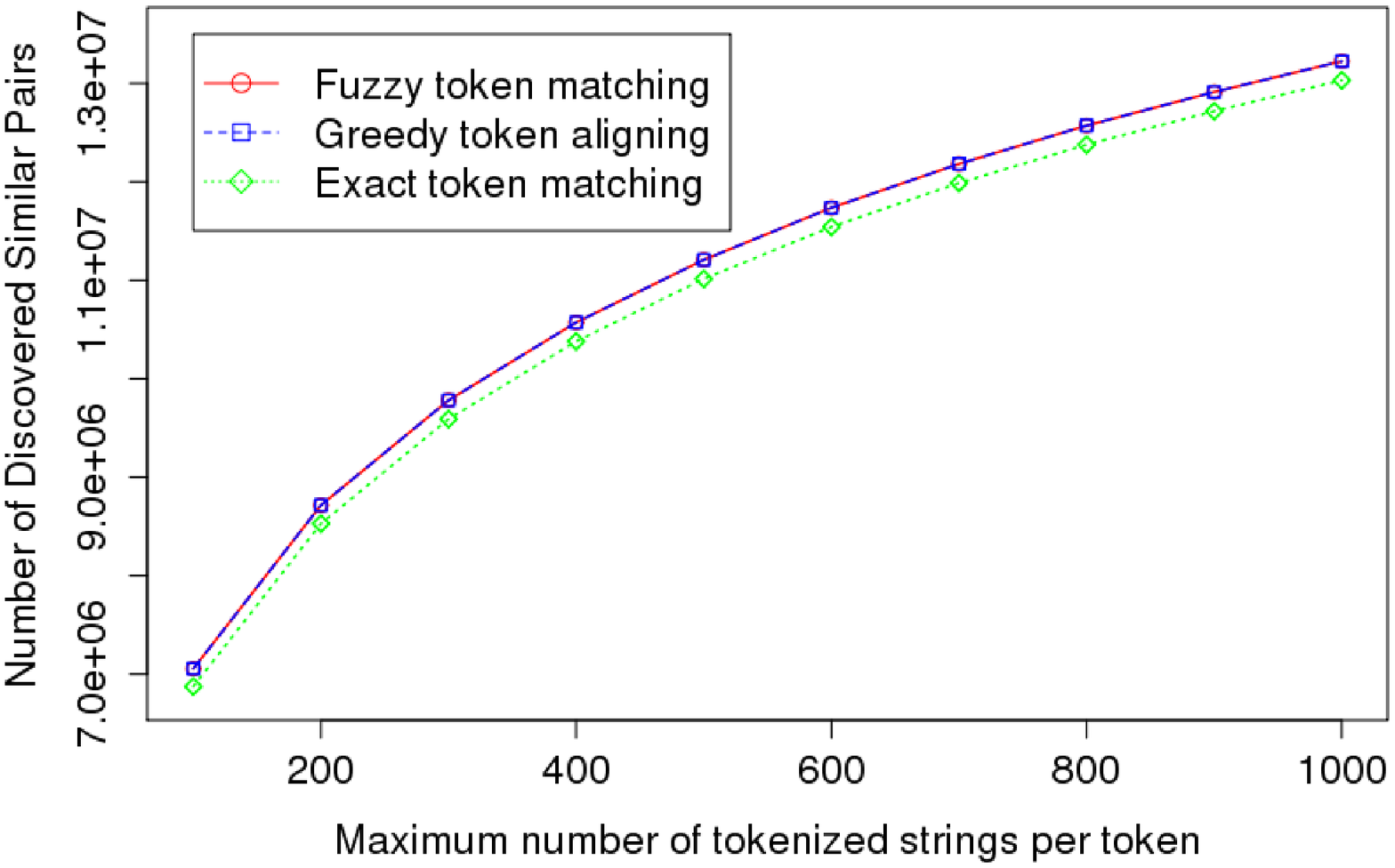}
%\vspace{-10pt}
\caption{Comparing the number of pairs of Tokenized-String Joiner (TSJ) while varying max-frequency ($M$) and the token matching and aligning algorithms.}
\label{fig:num_pairs_vs_max_freq}
%\end{minipage}
%\vspace{-10pt}
\end{figure}

The proposed approximations make TSJ err on the false negative side, guaranteeing the precision (the percentage of the discovered pairs that are truly similar) to be always $1.0$. However, the recall (the ratio between the number of the discovered pairs to the number of pairs discovered by fuzzy-token-matching) is not necessarily as perfect. To assess the impact of the proposed approximations on the recall, the number of pairs of tokenized strings that were found similar were reported while varying the two main parameters, $T$ and $M$. The numbers of similar pairs are plotted in \Figure~\ref{fig:num_pairs_vs_nsld} and \Figure~\ref{fig:num_pairs_vs_max_freq}, respectively.

\Figure~\ref{fig:num_pairs_vs_nsld} shows the number of discovered pairs while varying $T$ from $0.025$ to $0.225$\hide{ at a fixed interval of $0.025$}. The recall of greedy-token-aligning (exact-token-matching) was $1.0$ when $T$ was $0.025$ and decreased to $0.99993$ ($0.86655$) as $T$ reached $0.225$.

\Figure~\ref{fig:num_pairs_vs_max_freq} shows the number of discovered pairs while varying $M$ from $100$ to $\mbox{1,000}$\hide{ at a fixed interval of $100$}. For all values of $M$, the recall of greedy-token-aligning was stable around $0.999999$, and between $0.974$ and $0.985$ for exact-token-matching.

In the space of possible ($T$, $M$) thresholds, some deductions could be made around the reasonable point of  $(0.1, \mbox{1,000})$ from \Figure~\ref{fig:num_pairs_vs_nsld} and \Figure~\ref{fig:num_pairs_vs_max_freq}. First, the number of similar pairs of tokenized strings increases more aggressively by increasing $T$ than by increasing $M$. Second, increasing $T$ has more impact on the recall of the approximations proposed in \Section~\ref{sec:variation} than $M$. Below we analyze why this is the case for each of the two approximations in turn.

As $T$ increases, there is more room for fuzzy-token-matching to behave differently from greedy-token-aligning. When aligning the tokens of any pair of tokenized strings, at least a pair of aligned tokens has its $NLD$ smaller than $T$ (Theorem~\ref{theorem:NSLD_to_NLD}). Hence, larger $T$ translates to more ways to align the tokens of the two tokenized strings, making greedy-token-aligning finding the optimal alignment less likely.

On the contrary, an increase in $M$ does not highlight the difference between fuzzy-token-matching and greedy-token-aligning. An increase in $M$ merely increases the number of tokens considered for generating candidate pairs of tokenized strings, which does not impact the token aligning.

The increase in $T$ leaves room for fuzzy-token-matching to behave differently from exact-token-matching in generating the candidate pairs of tokenized strings. For any pair of tokenized strings to have their $NSLD$ below the $T$, at least two tokens have to exist, one from each tokenized string, such that their $NLD$ is below $T$ (Theorem~\ref{theorem:NSLD_to_NLD}). For small $T$ and relatively short name strings, small $NLD$ increases the chance of an exact match between these two tokens, which makes fuzzy-token-matching degenerate to exact-token-matching.

On the other hand, the increase in $M$ highlights the difference between fuzzy-token-matching and greedy-token-aligning to a lesser degree. Any similar pair of tokenized strings either share a token, or have a pair of tokens that can be fuzzily matched together. In the first case, exact-token-matching behave exactly like fuzzy-token-matching. In the second case, only fuzzy-token-matching discovers such candidate pairs. Since as $M$ increases, the tokens that become considered for discovering candidate pairs are popular, the first case becomes more prevalent.

%\vspace{-2pt}

\subsection{Lessons from Advertising Fraud Rings}

%\vspace{-2pt}

The TSJ framework was very effective in catching fraud rings. Using multiple string signals, TSJ allowed for discovering several fraud rings that were not detectable previously. Based on the evaluation above that was done on real full names, we strongly recommend using greedy-token-aligning for all values of $T$, and $M$. This results in almost no loss in recall, and notable enhancement in the runtime. We also recommend implementing grouping-on-one-string since it resulted in improving the runtime for all the experiments.

Using exact-token-matching when $T$ is set to a small or a moderate value (between $0.025$ and $0.1$) results in very minor loss in recall, with very significant improvement in the runtime. However, It is worth noting that the exact-token-matching approximation caught mainly the unsophisticated attacks that had the least monetary impact. The more sophisticated attacks were caught only by the fuzzy-token-matching algorithm. Hence, we recommend this approximation only for data integration and cleaning where missing some similar records does not have a significant financial impact, and the computational resources are scarce.

%\vspace{-2pt}

\subsection{Comparing the Accuracy of the Distance Measures}

%\vspace{-2pt}

The distances produced by $NSLD$ and the state-of-the-art distance measures are compared. The comparison was with the weighted versions of the set-based fuzzy similarity measures, $FJaccard$, $FCosine$, and $FDice$ \cite{WLF14}. For all the set-based fuzzy measures, the distance is taken as $1 - similarity$.

\hide{
Given the simple strings $s_{1} =$ ``{\tt chan kalan}'', and $s_{2} =$ ``{\tt chank alan}'', $NSLD(s_{1}^{t}, s_{2}^{t}) = 0.2$, while $1 - FJaccard(s_{1}^{t}, s_{2}^{t}) = 0.33$. When $s_{1} =$ ``{\tt alex thompson}'', and $s_{2} =$ ``{\tt alexa thomson}'', $NSLD(s_{1}^{t}, s_{2}^{t}) = 0.15$, while $1 - FJaccard(s_{1}^{t}, s_{2}^{t}) = 0.28$. When compared to $FJaccard$, $NSLD$ is more intuitive in the sense that it better represents the ratio of the characters edited to the length of the strings.
}

\begin{figure}
%\begin{minipage}{\smallfigwidth}
\centering
\epsfig{width = \smallfigwidth, figure = 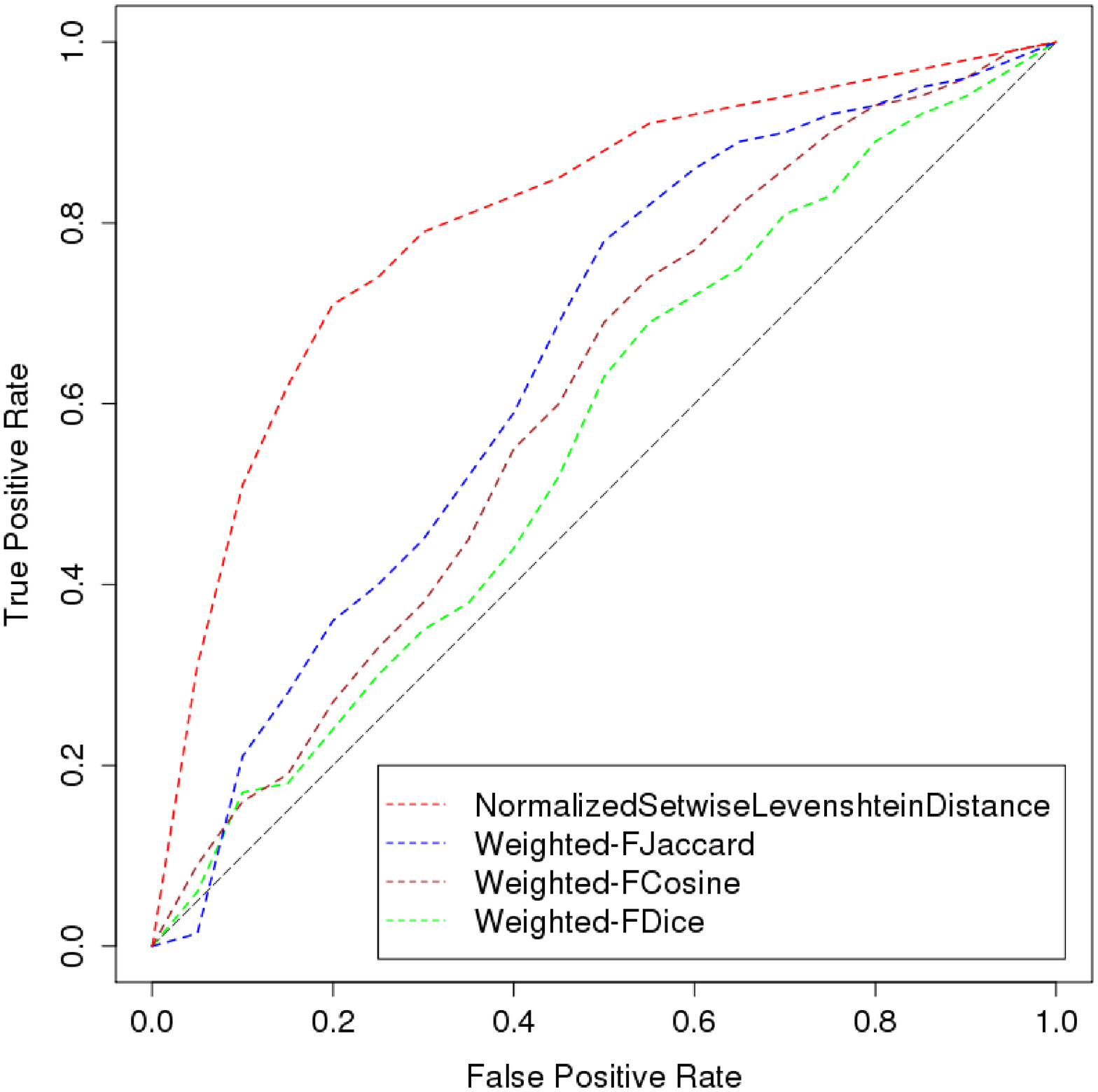}
%\vspace{-10pt}
\caption{The ROC curves of NSLD, wighted FJaccard, weighted FCosine, and weighted FDice when predicting fraudulent accounts based on the distance between the old and new names on an account.}
\label{fig:nsld_vs_fjaccard_vs_fcosine_vs_fdice}
%\end{minipage}
%\vspace{-10pt}
\end{figure}

We could not evaluate the quality of the set-based fuzzy similarity measures on our dataset, since the join algorithm proposed in \cite{WLF14} is serial and does not scale to large-scale datasets. However, we assessed how $NSLD$ and the set-based fuzzy measures can be used as proxies to assess if an account is fraudulent or not. We have selected a sample of $\mbox{10,000}$ accounts that have their names changed. Half the sample were known legitimate accounts and the other half were known fraudulent accounts. The name changes on the legitimate accounts happen in rare cases, such as legal name changes, or name abbreviation, e.g., from ``{\tt William}'' to ``{\tt Bill}''. On the other hand, name changes on fraudulent accounts are usually very drastic, since, attackers who specialize in account creation are not those who specialize in account exploitation \cite{TMGKP13}. The account-creation attacker typically chooses a random name. When the credentials are sold to the attacker, the account name is drastically changed. The $NSLD$ and the set-based fuzzy measures were used to measure the distance between the old and the new account names. The ROC curves are in \Figure~\ref{fig:nsld_vs_fjaccard_vs_fcosine_vs_fdice}.

Assuming the correlation between the magnitude of the name change and the likelihood of fraud, $NSLD$ is superior to all these set-based fuzzy measures when quantifying the distances between names on accounts. The existing tokenized-string measures assume the edits are mostly non-malicious and are either manual typos or OCR errors. This assumption does not apply when an adversary strives to game the measures.

%\vspace{-2pt}

\subsection{Comparing TSJ with the State of the Art Join Algorithms}

%\vspace{-2pt}

TSJ is the first distributed similarity-join algorithm for tokenized strings. However, the existing distributed metric-space join algorithms lend themselves to a comparison with TSJ, since the proposed distance, $NSLD$, is a metric.

\begin{figure}
%\begin{minipage}{\smallfigwidth}
\centering
\epsfig{width = \smallfigwidth, figure = 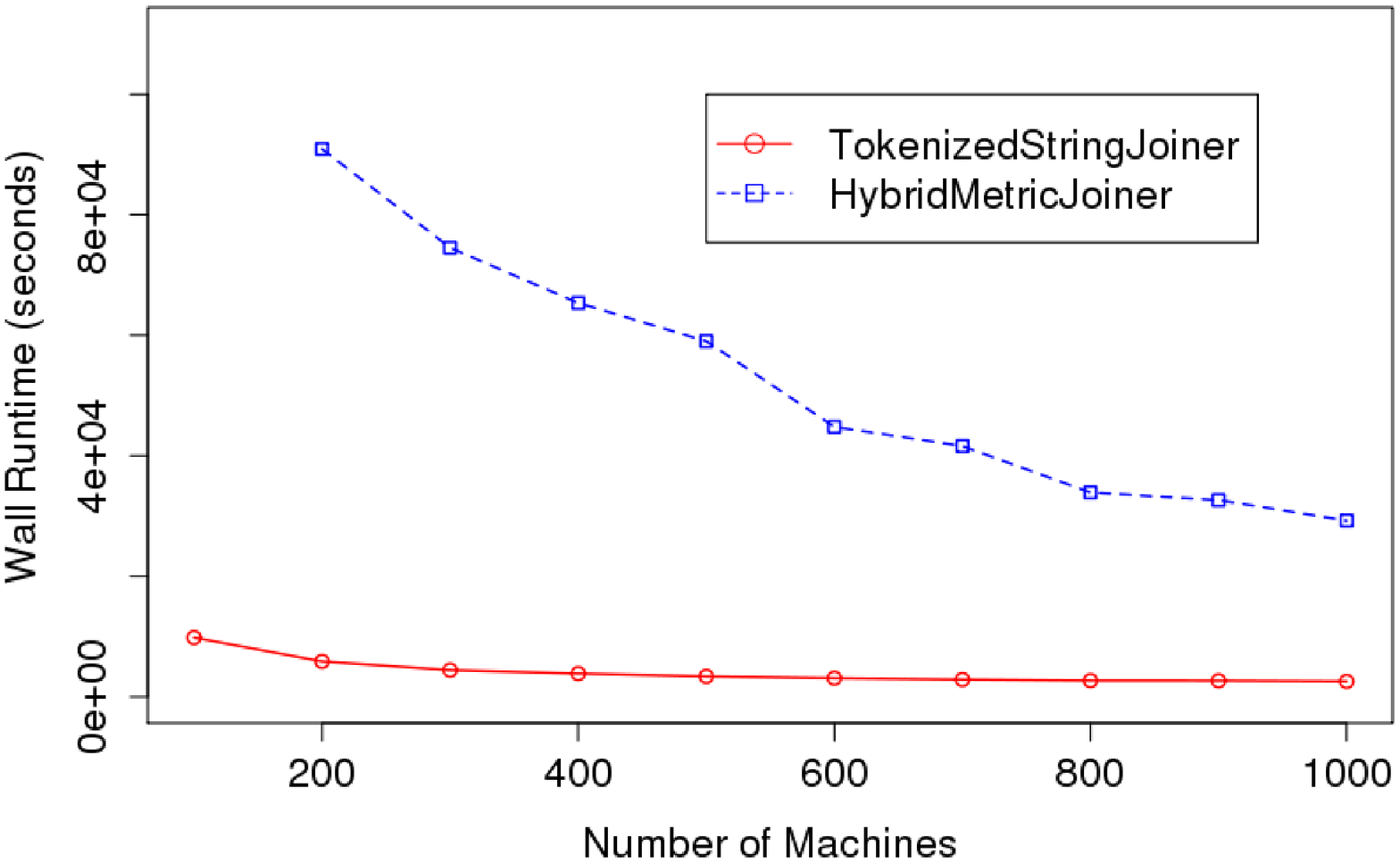}
%\vspace{-10pt}
\caption{Comparing the runtime of Tokenized-String Joiner (TSJ) and the Hybrid Metric Joiner (HMJ) while varying the MapReduce machines.}
\label{fig:run_time_vs_num_machines_deduping_tsj_vs_hybrid}
%\end{minipage}
%\vspace{-10pt}
\end{figure}

We compared TSJ to a state-of-the-art metric-space join algorithm. The Hybrid Metric Joiner (HMJ) is an in-house-built algorithm that is hybrid of the most scalable and efficient algorithms \cite{WMP13,DHC14} proposed for metric-space joins and reviewed in \Section~\ref{sec:related}. As proposed in \cite{DHC14}, the tokenized strings are dissected into partitions among centroids by dissecting the space among the centroids using Voronoi hyperplanes, and are assigned to neighboring partitions using the general filter in \cite{DHC14}. The symmetry of the distance metric was exploited to reduce comparing tokenized strings from neighboring partitions, as proposed in \cite{WMP13}. Each partition is recursively repartitioned either using sub-centroids as proposed in \cite{WMP13}, or using a $2$-dimensional grid as proposed in \cite{DHC14}, depending on how the tokenized strings are scattered within the partition. Finally, the computation is made more efficient by exploiting the triangle inequality to output cliques and bicliques of tokenized strings as described in \cite{WMP13}. TSJ and HMJ were run while varying the number of machines from $100$ to $\mbox{1,000}$\hide{ at a fixed interval of $100$}, and the runtimes are plotted in \Figure~\ref{fig:run_time_vs_num_machines_deduping_tsj_vs_hybrid}. HMJ did not finish on $100$ machines in a reasonable amount of time. For all the other configs, TJS was $12$ to $15$ times faster than HMJ.

The metric-space techniques work best when the data is fairly distributed in the space. However, when visualizing the tokenized strings as multi-dimensional points in a metric space, they form a large number of fairly dense clusters, due to sharing tokens. This results in load imbalance between the workers. TJS on the other hand transforms the join problem to the token domain, performs a join using a very specialized and scalable string-join algorithm, and uses the results to generate candidate pairs of tokenized strings. Since most of the work happens in the token domain, the poorly distributed tokenized strings do not impact the performance.

\section{Conclusion}
Existing tokenized-string distance measures are sensitive to the order of the tokens, asymmetric, or are hard to tune since they require setting multiple independent thresholds. Moreover, their join algorithms are serial and hence unscalable.

In this paper, we motivated and introduced a novel distance metric between tokenized strings, Normalized Setwise Levenshtein Distance ($NSLD$). To the best of our knowledge, $NSLD$ is the first metric distance for tokenized strings. Moreover, $NSLD$ is the only distance measure that guarantees carrying the distance threshold from the tokenized strings to their tokens. This allows for transforming the join from the tokenized-string domain to the tokens domain. The tokens domain is more manageable, since the number of distinct tokens is typically orders of magnitude smaller than that of distinct tokenized strings, and the literature of string-join algorithm is richer. Based on this property, we propose a very specialized and scalable framework, Tokenized-String Joiner (TSJ), which adopts existing string-join algorithms as building blocks to perform $NSLD$-based joins. The scalability of the proposed framework, and the effectiveness and accuracy of the proposed approximations are established by our evaluation on tens of millions of names on {\Google} accounts. We demonstrated the superiority of $NSLD$ over the state-of-the-art weighted set-based fuzzy similarity measures in terms of accuracy, and the superiority of the tokenized-string-specific TSJ over the metric-spaces joins in terms of scalability and efficiency.
\hide{
Our future work spans multiple directions. The most important and potentially rewarding is leveraging the proved properties of $NLD$ and $NSLD$ for further pruning of pairs of candidate tokenized strings.
}

\balance
\bibliographystyle{abbrv}
\scriptsize
\footnotesize
%\small
%\tiny

\bibliography{strings}

\appendix
\section{Appendix}

%Below are the theorem proofs.

\subsection{Proof of Lemma~\ref{lemma:NLD_threshold}}
\label{lemma_proof:NLD_threshold}

Since $|y| \geq |x|$, and from \Definition~\ref{definition:NLD}, $NLD(x, y) = \frac{2}{\frac{|x|+|y|}{LD(x, y)} + 1}$, then $|y| - |x| \leq LD(x, y) \leq |y|$.

\subsection{Proof of Lemma~\ref{lemma:SLD_metric}}
\label{lemma_proof:SLD_metric}

Trivially, $SLD(x^{t}, x^{t}) = 0$. Moreover, $SLD(x^{t}, y^{t}) = SLD(y^{t}, x^{t}) \forall x^{t}, y^{t}$. To prove the Triangle Inequality, let $x^{t} = \{x^{t1}, x^{t2}, \dots, x^{tm}\}$, $y^{t} = \{y^{t1}, y^{t2}, \dots, y^{tn}\}$, and $z^{t} = \{z^{t1}, z^{t2}, \dots, z^{to}\}$. Let $k = max\{m, n, o\}$. Construct three auxiliary tokenized strings $x^{t \prime} = \{x^{t1}, x^{t2}, \dots, x^{tk}\}$, $y^{t \prime} = \{y^{t1}, y^{t2}, \dots, y^{tk}\}$, and $z^{t \prime} = \{z^{t1}, z^{t2}, \dots, z^{tk}\}$, where extra tokens are empty strings. Clearly, $SLD(x^{t}, y^{t}) = SLD(x^{t \prime}, y^{t \prime})$, $SLD(y^{t}, z^{t}) = SLD(y^{t \prime}, z^{t \prime})$, and $SLD(x^{t}, z^{t}) = SLD(x^{t \prime}, z^{t \prime})$. There is a transforming path for each token tuple $x^{tx_{i}} \to y^{ty_{i}} \to z^{tz_{i}}, \forall i \in [1, k]$. From the Triangle Inequality of $LD(\cdot, \cdot)$, it follows that $LD(x^{tx_{i}}, y^{ty_{i}}) + LD(y^{ty_{i}}, z^{tz_{i}}) \geq LD(x^{tx_{i}}, z^{tz_{i}})$. Summing on all $k$ tokens, $\sum_{i}^{k}LD(x^{tx_{i}}, y^{ty_{i}}) + \sum_{i}^{k}LD(y^{ty_{i}}, z^{tz_{i}}) \geq \sum_{i}^{k}LD(x^{tx_{i}}, z^{tz_{i}})$. Hence, $SLD(x^{t}, y^{t}) + SLD(y^{t}, z^{t}) \geq SLD(x^{t}, z^{t})$.

\subsection{Proof of Lemma~\ref{lemma:NSLD_threshold_1}}
\label{lemma_proof:NSLD_threshold_1}
Trivially, $NSLD(x^{t}, x^{t}) = 0$. In the other extreme, let $x^t$ only be empty. $\mathcal{L}(x^{t}) = 0$. From \Definition~\ref{definition:SLD}, $SLD(x^{t}, y^{t}) = \mathcal{L}(y^{t})$. Hence, $NSLD(x^{t}, y^{t}) = \frac{2 \times \mathcal{L}(y^{t})}{2 \times \mathcal{L}(y^{t})} = 1$.

\subsection{Proof of Lemma~\ref{lemma:NSLD_threshold_2}}
\label{lemma_proof:NSLD_threshold_2}
 $\mathcal{L}(y^{t}) \geq \mathcal{L}(x^{t})$ is assumed. From \Definition~\ref{definition:NSLD}, $NSLD(x^{t}, y^{t}) = \frac{2}{\frac{\mathcal{L}(x^{t}) + \mathcal{L}(y^{t})}{SLD(x^{t}, y^{t})} + 1}$. It follows that $\mathcal{L}(y^{t}) - \mathcal{L}(x^{t})$ $\leq$ $SLD(x^{t}, y^{t})$ $\leq$ $\mathcal{L}(y^{t})$.

\subsection{Proof of Theorem~\ref{theorem:NSLD_metric}}
\label{theorem_proof:NSLD_metric}

Trivially, $NSLD(x^{t}, x^{t}) = 0$. Moreover, $NSLD(x^{t}, y^{t}) = NSLD(y^{t}, x^{t}) \forall x^{t}, y^{t}$. To prove the Triangle Inequality, let $x^{t} = \{x^{t1}, x^{t2}, \dots, x^{tm}\}$, $y^{t} = \{y^{t1}, y^{t2}, \dots, y^{tn}\}$, and similarly $z^{t} = \{z^{t1}, z^{t2}, \dots, z^{to}\}$. we prove the case $NSLD(x^{t}, y^{t}) + NSLD(y^{t}, z^{t}) \geq NSLD(x^{t}, z^{t})$.

Because of Lemma~\ref{lemma:NSLD_threshold_1}, if any of $x^{t}$, $y^{t}$, or $z^{t}$ are empty, the Triangle Inequality is trivially satisfied. Thus, we consider the case where none of the three tokenized strings are empty.

Again, because of Lemma~\ref{lemma:NSLD_threshold_1}, if $NSLD(x^{t}, y^{t}) \geq NSLD(x^{t}, z^{t})$ or $NSLD(y^{t}, z^{t}) \geq NSLD(x^{t}, z^{t})$, then the Triangle Inequality is trivially satisfied. Thus, we consider the case where $NSLD(x^{t}, z^{t})$ is greater than both $NSLD(y^{t}, z^{t})$ and $NSLD(x^{t}, y^{t})$.

\begin{align}
\label{eq:NSLD_metric_basic_condition}
NSLD(x^{t}, y^{t}) &< NSLD(x^{t}, z^{t}), and \nonumber \\
NSLD(y^{t}, z^{t}) &< NSLD(x^{t}, z^{t}).
\end{align}

From \Relation~\ref{eq:NSLD_metric_basic_condition}, we further derive the following relations.

\begin{align}
\label{eq:NSLD_metric_basic_condition_rewritten}
\left(
\frac{NSLD(x^{t}, z^{t})}{2 - NSLD(x^{t}, z^{t})} - \frac{NSLD(x^{t}, y^{t})}{2 - NSLD(x^{t}, y^{t})}
\right) &> 0, and \nonumber \\
\left(\frac{NSLD(x^{t}, z^{t})}{2 - NSLD(x^{t}, z^{t})} - \frac{NSLD(y^{t}, z^{t})}{2 - NSLD(y^{t}, z^{t})}
\right) &> 0.
\end{align}

Lemma~\ref{lemma:SLD_metric}, the Triangle Inequality of $SLD$, gives the following relations.

\begin{equation}
\label{eq:NSLD_metric_SLD}
SLD(x^{t}, y^{t}) + SLD(y^{t}, z^{t}) \geq SLD(x^{t}, z^{t})
\end{equation}

From \Definition~\ref{definition:NSLD}, we express $SLD$ in terms of $NSLD$.

\begin{align}
\label{eq:NSLD_metric_SLD_in_terms_of_NSLD}
SLD(x^{t}, y^{t}) &= \frac{NSLD(x^{t}, y^{t})}{2 - NSLD(x^{t}, y^{t})} \times
\left(
\mathcal{L}(x^{t}) + \mathcal{L}(y^{t})
\right), \nonumber \\
SLD(y^{t}, z^{t}) &= \frac{NSLD(y^{t}, z^{t})}{2 - NSLD(y^{t}, z^{t})} \times
\left(
\mathcal{L}(y^{t}) + \mathcal{L}(z^{t})
\right), and \nonumber \\
SLD(x^{t}, z^{t}) &= \frac{NSLD(x^{t}, z^{t})}{2 - NSLD(x^{t}, z^{t})} \times
\left(
\mathcal{L}(x^{t}) + \mathcal{L}(z^{t})
\right).
\end{align}

Combining \Relations~\ref{eq:NSLD_metric_SLD} and~\ref{eq:NSLD_metric_SLD_in_terms_of_NSLD}, we get the following relations.

\begin{multline*}
\frac{NSLD(x^{t}, y^{t})}{2 - NSLD(x^{t}, y^{t})} \times
\left(\mathcal{L}(x^{t}) + \mathcal{L}(y^{t})
\right) + \\
\frac{NSLD(y^{t}, z^{t})}{2 - NSLD(y^{t}, z^{t})} \times
\left(
\mathcal{L}(y^{t}) + \mathcal{L}(z^{t})
\right) \geq \\
\frac{NSLD(x^{t}, z^{t})}{2 - NSLD(x^{t}, z^{t})} \times
\left(
\mathcal{L}(x^{t}) + \mathcal{L}(z^{t})
\right)
\end{multline*}

This can be rewritten as follows.

\begin{multline}
\label{eq:NSLD_metric_SLD_in_terms_of_NSLD_expanded}
\left(
\frac{NSLD(x^{t}, y^{t})}{2 - NSLD(x^{t}, y^{t})} +
\frac{NSLD(y^{t}, z^{t})}{2 - NSLD(y^{t}, z^{t})}
\right) \times \mathcal{L}(y^{t}) \geq \\
\left(
\frac{NSLD(x^{t}, z^{t})}{2 - NSLD(x^{t}, z^{t})} - 
\frac{NSLD(x^{t}, y^{t})}{2 - NSLD(x^{t}, y^{t})}
\right) \times \mathcal{L}(x^{t}) + \\
\left(
\frac{NSLD(x^{t}, z^{t})}{2 - NSLD(x^{t}, z^{t})} -
\frac{NSLD(y^{t}, z^{t})}{2 - NSLD(y^{t}, z^{t})}
\right) \times \mathcal{L}(z^{t})
\end{multline}

From both Lemmas~\ref{lemma:NSLD_threshold_1} and~\ref{lemma:NSLD_threshold_2}, $1 - \frac{\mathcal{L}(x^{t})}{\mathcal{L}(y^{t})} \leq NSLD(x^{t}, y^{t}) \forall x^{t}, y^{t}$. This can be rewritten as the following two relations.

\begin{equation}
\label{eq:NSLD_metric_lemmas_rewritten_x}
\mathcal{L}(x^{t}) \geq (1 - NSLD(x^{t}, y^{t})) \times \mathcal{L}(y^{t}) \forall x^{t}, y^{t}
\end{equation}

\begin{equation}
\label{eq:NSLD_metric_lemmas_rewritten_y}
\mathcal{L}(z^{t}) \geq (1 - NSLD(y^{t}, z^{t})) \times \mathcal{L}(y^{t}) \forall y^{t}, z^{t}
\end{equation}

Putting together \Relations~\ref{eq:NSLD_metric_basic_condition_rewritten},~\ref{eq:NSLD_metric_SLD_in_terms_of_NSLD_expanded},~\ref{eq:NSLD_metric_lemmas_rewritten_x}, and~\ref{eq:NSLD_metric_lemmas_rewritten_y} yields the following.

\begin{equation}
\label{eq:NSLD_metric_final}
\begin{split}
\left(
\frac{NSLD(x^{t}, y^{t})}{2 - NSLD(x^{t}, y^{t})} +
\frac{NSLD(y^{t}, z^{t})}{2 - NSLD(y^{t}, z^{t})}
\right) \times \mathcal{L}(y^{t}) & \geq \\
\left(
\frac{NSLD(x^{t}, z^{t})}{2 - NSLD(x^{t}, z^{t})} - 
\frac{NSLD(x^{t}, y^{t})}{2 - NSLD(x^{t}, y^{t})}
\right) \times
\mathcal{L}(y^{t})
& \times \\
\left(
1 - NSLD(x^{t}, y^{t})
\right) & + \\
\left(
\frac{NSLD(x^{t}, z^{t})}{2 - NSLD(x^{t}, z^{t})} -
\frac{NSLD(y^{t}, z^{t})}{2 - NSLD(y^{t}, z^{t})}
\right) \times
\mathcal{L}(y^{t})
& \times \\
\left(
1 - NSLD(y^{t}, z^{t})
\right)
\end{split}
\end{equation}

Since $\mathcal{L}(y^{t}) > 0$, then \Relation~\ref{eq:NSLD_metric_final} can be simplified as follows.

\begin{equation*}
\begin{split}
\left(
\frac{NSLD(x^{t}, y^{t})}{2 - NSLD(x^{t}, y^{t})}
\right) & \times
\left(
1 +
\left(1 - NSLD(x^{t}, y^{t})
\right)
\right) + \\
\left(
\frac{NSLD(y^{t}, z^{t})}{2 - NSLD(y^{t}, z^{t})}
\right) & \times
\left(
1 +
\left(
1 - NSLD(y^{t}, z^{t})
\right)
\right) \geq \\
\left(
\frac{NSLD(x^{t}, z^{t})}{2 - NSLD(x^{t}, z^{t})}
\right) & \times
\left(
\left(
1 - NSLD(x^{t}, y^{t})
\right) +
\left(
1 - NSLD(y^{t}, z^{t})
\right)
\right)
\end{split}
\end{equation*}

This can be further simplified as follows.

\begin{multline}
\label{eq:NSLD_metric_final_Ly_greater_than_0}
\frac{\left(NSLD(x^{t}, y^{t}) + NSLD(y^{t}, z^{t})\right)}{NSLD(x^{t}, z^{t})} \geq \\
\frac{2 - \left(NSLD(x^{t}, y^{t}) + NSLD(y^{t}, z^{t})\right)}{2 - NSLD(x^{t}, z^{t})}
\end{multline}

\Relation~\ref{eq:NSLD_metric_final_Ly_greater_than_0} implies $NSLD(x^{t}, y^{t}) + NSLD(y^{t}, z^{t}) \geq NSLD(x^{t}, z^{t})$.

\subsection{Proof of Theorem~\ref{theorem:NSLD_to_NLD}}
\label{theorem_proof:NSLD_to_NLD}

The proof is by contradiction. W.L.O.G, let $NSLD$$(x^{t}$, $y^{t})$ $\leq T$; $x^{t}$ and $y^{t}$ has two tokens each, $x^{t1}, x^{t2}$ and $y^{t1}, y^{t2}$, respectively; and to achieve minimum $SLD$, $x^{ti}$ is transformed into $y^{ti}$, $\forall i \in \{1,2\}$. Assuming $NLD(x^{ti}, y^{ti}) > T$, $\forall i \in \{1,2\}$, it follows that $\frac{2 \times LD(x^{ti}, y^{ti})}{|x^{ti}| + |y^{ti}| + LD(x^{ti}, y^{ti})} > T$, $\forall i \in \{1,2\}$. Let $f_{1} = 2 \times LD(x^{t1}, y^{t1})$, $f_{2} = |x^{t1}| + |y^{t1}| + LD(x^{t1}, y^{t1})$, $f_{3} = 2 \times LD(x^{t2}, y^{t2})$, $f_{4} = |x^{t2}| + |y^{t2}| + LD(x^{t2}, y^{t2})$. Hence, $\frac{f_{1}}{f_{2}} > T$ and $\frac{f_{3}}{f_{4}} > T$. Notice that $NSLD(x^{t}, y^{t}) = \frac{f_{1} + f_{3}}{f_{2} + f_{4}}$, and that the initial assumption was $NSLD(x^{t}, y^{t}) \leq T$, which contradicts $\frac{f_{1}}{f_{2}} > T$ and $\frac{f_{3}}{f_{4}} > T$. The argument is generalizable for any $\mathcal{T}(x^{t})$ and $\mathcal{T}(y^{t})$.

\subsection{Proof of Lemma~\ref{lemma:NLD_to_LD_threshold}}
\label{lemma_proof:NLD_to_LD_threshold}

From \Definition~\ref{definition:NLD}, $\frac{2 \times LD(x, y)}{|x| + |y| + LD(x, y)} \leq T$, which implies $LD(x, y) \leq \frac{T \times (|x| + |y|)}{2 - T}$.
If $|x| \leq |y|$, substituting with the upper bound of $|x|$, which is $|y|$, and taking the lower bound of the r.h.s. since $LD(x, y)$ is a whole number, yields $LD(x, y) \leq \lfloor \frac{2 \times T \times |y|}{2 - T} \rfloor$.
If $|x| > |y|$, then $|x| \leq LD(x, y) + |y|$. Substituting with the upper bound of $|x|$ in \Definition~\ref{definition:NLD} yields $\frac{2 \times LD(x, y)}{|y| + LD(x, y) + |y| + LD(x, y)} \leq T$. This can be simplified to $LD(x, y) \leq \lfloor \frac{T \times |y|}{1-T} \rfloor$.

\subsection{Proof of Lemma~\ref{lemma:lowerbound_for_x}}
\label{lemma_proof:lowerbound_for_x}

From \Definition~\ref{definition:NLD}, $\frac{2 \times LD(x, y)}{|x| + |y| + LD(x, y)} \leq T$, which implies $LD(x, y) \leq \frac{T \times (|x| + |y|)}{2 - T}$. Substituting with the lower bound of $LD(x, y)$, which is $|y| - |x|$, yields $|y| - |x| \leq \frac{T \times (|x| + |y|)}{2 - T}$. This can be simplified as $2 \times |y| - 2 \times |x| \leq 2 \times T \times |y|$, and further simplified as $|y| \times (1 - T) \leq |x|$. Taking the upper bound of the l.h.s. since $|x|$ has to be a whole number establishes the relationship in the lemma.

\subsection{Proof of Lemma~\ref{lemma:histogram_NLD_to_LD_threshold}}
\label{lemma_proof:histogram_NLD_to_LD_threshold}

From \Definition~\ref{definition:NLD}, $\frac{2 \times LD(x, y)}{|x| + |y| + LD(x, y)} > T$, which implies $LD(x, y) > \frac{T \times (|x| + |y|)}{2 - T}$.
If $|x| \leq |y|$, substituting with the lower bound of $|x|$, which is $0$, and taking the lower bound of the r.h.s. since $LD(x, y)$ is a whole number, yields $LD(x, y) > \lfloor \frac{T \times |y|}{2 - T} \rfloor$.
If $|x| > |y|$, substituting with the lower bound of $|x|$, which is $|y|$, and taking the lower bound of the r.h.s. since $LD(x, y)$ is a whole number, yields $LD(x, y) > \lfloor \frac{2 \times T \times |y|}{2-T} \rfloor$.

\end{document}